\documentclass[11pt, a4paper, oneside]{article}
\usepackage{inputenc}
\usepackage[section]{placeins}

\usepackage[maxbibnames=10, backend=biber, autocite=superscript, natbib, backend=bibtex, bibstyle=authoryear, giveninits=true,
style=authoryear,maxcitenames=2]{biblatex}%style=authoryear,maxcitenames=2citestyle=numeric

\usepackage[T1]{fontenc}
\usepackage{paralist}
\usepackage{setspace}
\usepackage{setspace}
\usepackage{graphicx}
\usepackage{graphics}
\usepackage{wrapfig}
\usepackage{amsmath}
\usepackage{tabularx}
\usepackage[dvipsnames]{xcolor}
\usepackage[margin=2cm]{geometry}
\usepackage{mathtools}
\usepackage{tcolorbox}
\usepackage{float}
\usepackage[normalem]{ulem}
\definecolor{arsenic}{rgb}{0.23, 0.27, 0.29}
\usepackage[font=small,skip=5pt]{caption}
\usepackage{titlesec}
\usepackage{arydshln}
\usepackage{soul}
\usepackage{subcaption}
\usepackage{adjustbox}
\usepackage{lineno}
\usepackage{setspace}
\usepackage{multirow}
\usepackage{multicol}

\usepackage{listings}
\usepackage{titling}
\usepackage{array}
\usepackage{makecell}
\usepackage{MnSymbol}
\usepackage{mathtools}
\usepackage{scrextend}
\usepackage{caption}
\usepackage{hyperref}
\usepackage{siunitx}
\DeclareSIUnit{\Molar}{\textsc{M}}
\DeclareCiteCommand{\citeyearpar}
    {}
    {\mkbibparens{\bibhyperref{\printdate}}}
    {\multicitedelim}
    {}

\usepackage{etoolbox}
\AtBeginEnvironment{tabular}{\small}

\begin{document}
%\setmainfont{Arial}

%\defaultfontfeatures{ Scale=MatchLowercase, Ligatures = TeX }
%\setmainfont{Arial}
%\setsansfont{Arial}
%\setmonofont{Andale Mono}
%\setmathfont{GFSNeohellenicMath.otf}
%\setmathfont[range=up]{Arial}
%\setmathfont[range=it]{Arial Italic}
%\setmathfont[range=bfup]{Arial Bold}
%\setmathfont[range=bfit]{Arial Bold Italic}
%\setmathfont[range=tt]{Andale Mono}
\bigskip
{\Large
\noindent A bibliometric study on mathematical oncology: interdisciplinarity, internationality, collaboration and trending topics}
%past and present trends in mathematical oncology
\begin{flushleft}
\bigskip 
Kira Pugh\textsuperscript{1$\circledast$}, 
Linnéa Gyllingberg\textsuperscript{2},
Stanislav Stratiev\textsuperscript{3},
Sara Hamis\textsuperscript{1$\circledast$}.

\bigskip
\textbf{1} Department of Information Technology, Uppsala University, Uppsala, Sweden. 
\\
\textbf{2} Department of Mathematics, University of California Los Angeles, CA, USA.
\\
\textbf{3} {Northell, Birmingham, UK}.
\\

\end{flushleft}

\smallskip
\noindent{\bf Author ORCiDs}: 
\smallskip
\textbf{KP:} 0000-0001-5588-7014; 
\textbf{LG:} 0000-0002-8745-4480;
\textbf{SH:} 0000-0002-1105-8078.

\bigskip
\noindent{\bf $\circledast$ Corresponding authors}: kira.pugh@it.uu.se, sara.hamis@it.uu.se. 

\thispagestyle{empty} 
 %\linenumbers
\vspace{5cm}
\section*{Abstract}

Mathematical oncology is an interdisciplinary research field where the mathematical sciences meet cancer research. 
Being situated at the intersection of these two fields makes mathematical oncology highly dynamic, as practicing researchers are incentivised to quickly adapt to both technical and medical research advances. 
Determining the scope of mathematical oncology is therefore not straightforward; however, it is important for purposes related to funding allocation, education, scientific communication, and community organisation. 
To address this issue, we here conduct a bibliometric analysis of mathematical oncology. 
We compare our results to the broader field of mathematical biology, and position our findings within theoretical science of science frameworks. 

Based on article metadata and citation flows, our results provide evidence that mathematical oncology has undergone a significant evolution since the 1960s marked by increased interactions with other disciplines, geographical expansion, larger research teams, and greater diversity in studied topics. 
The latter finding contributes to the greater discussion on which models different research communities consider to be valuable in the era of big data and machine learning. 
Further, the results presented in this study quantitatively motivate 
that international collaboration networks should be supported to enable new countries to enter and remain in the field, and that 
mathematical oncology benefits both mathematics and the life sciences. 

\bigskip
\noindent{\bf Keywords}: Mathematical cancer research, mathematical biology, bibliometric analysis, science of science, scientometrics. 
\newpage 
\clearpage
\pagenumbering{arabic} 
\section{Introduction}  
\vspace{-.2cm}

Mathematical oncology is an interdisciplinary research field in which mathematical modelling, analysis, and simulations are used to study cancer \citep{Altrock2015}. 
Since the term {\it mathematical oncology} was introduced in the literature in the early 2000s \citep{Gatenby2003}, 
the application areas of mathematics in cancer research have increased with the recent surge in available cancer data ranging from 
omics data on the gene expression level \citep{Vandereyken2023}, 
to imaging data on the cell and tumour levels \citep{Bond2022,Rong2024}, 
biomarker data on the individual patient level \citep{Passaro2024}, 
and large-scale register data on the human population level \citep{Tucker2019}. 

Therefore, although mathematical oncology is not a new research field, with early models of tumour growth in mice published in the 1930s \citep{Mayneord1932}, it is dynamic. The field's objective to study one of our biggest health threats -- cancer -- incentivises researchers to quickly adapt to advances pertaining to new cancer data, therapies, and clinical practices. 
Such advances have the ability to rapidly alter global research directions, as opposed to mathematical research advances that promote methodological revolutions that slowly change scientific practices \citep{KRAUSS2024e36066}. 
The fact that mathematical oncology is motivated by cancer research, yet grounded in mathematics, 
provides a tangible example of how
research fields, particularly interdisciplinary ones, are not rigidly defined \citep{McGillivray2022}.
Rather, they change over time which complicates the practice of clearly defining the delimiters that describe what a specific research field is, and what it is not. 

The ability to quantitatively understand and describe a research field is helpful for actors operating both within and outside the field, including researchers, students, policy-makers, stake-holders, and the general public.
Importantly, such an understanding can influence the design of university and school curricula, resource allocation within funding bodies and institutions, and science communication strategies.

One effective approach to quantitatively study research fields is through bibliometric analysis, where statistical methods are used to analyse patterns in science communication data. 
Bibliometric studies enable large-scale quantitative mappings of research fields, including analysis of historical, current, and emerging research trends \citep{Donthu2021,otavio2019}. 
Platforms that host large databases of scientific publication data, such as Web of Science \citep{webofscience}, Scopus \citep{scopus}, and OpenAlex \citep{OpenAlex}, facilitate bibliometric data collection. Coupled with the uptick in user-friendly bibliometric visualisation tools, including VOSviewer \citep{van2009vosviewer}, biblioshiny \citep{aria2017bibliometrix}, CiteSpace \citep{chen2006citespace}, and CitNetExplorer \citep{van2014citnetexplorer}, it is becoming increasingly accessible to perform bibliometric studies, as is reflected by their rapid growth in numbers \citep{Cheng2024}. 

Today, it is common practice for research councils and institutes to perform bibliometric data analysis to evaluate the impact of published works produced by specified researchers and projects \citep{Thelwall2023}. 
Others who conduct bibliometric work include both science-of-science researchers, potentially ``looking in'' on a research field from the outside, and practitioners within the studied field. 
When coupling domain knowledge with theoretical frameworks from science-of-science, a broad range of topics that are traditionally studied qualitatively can be quantitatively assessed through bibliometric analysis \citep{Lund2021}. 
These topics include research ethics \citep{Gureyev2022}, values \citep{birhane2022values} and gender equality \citep{Brck2023}.  

Of relevance to this article, previous bibliometric studies have analysed the intersection of cancer research and disciplines that neighbour mathematics such as artificial intelligence \citep{koccak2024development}, machine learning \citep{Lin2023}, and deep learning \citep{Wang2024}. These studies largely focus on comparing publication and citation activity between countries and institutions, and highlight the rise of machine learning methods following the last decade's string of machine learning breakthroughs that received attention from both the scientific and broader community.

In this work, we contribute a bibliometric study on the intersection of cancer research and mathematics, {\em i.e.,} {\it mathematical oncology}. 
By analysing article metadata, we investigate interdisciplinarity, internationality, and collaboration patterns, as well as trending research topics in the field, between the years 1961 and 2024. 
We quantitatively compare our results to trends in the more general field {\it mathematical biology}, 
and position our findings within theoretical frameworks rooted in science of science \citep{fortunato2018science} and culture of science \citep{Franklin1995}.
\vspace{-.1cm}

\section{Methods}
\vspace{-.1cm}

\subsection{Study aims}
The design of a bibliometric study should be tailored to its aims \citep{oztrk2024}. 
In this study, our aim is to answer the following core research questions on interdisciplinarity, internationality, collaboration, and trending topics, respectively:
\begin{enumerate}[R1.]
   \item Which disciplines have influenced, and are 
   influenced by, mathematical oncology? In other words, who is mathematical oncology for? 
   \item How has mathematical oncology expanded globally over time? Do new countries enter the field independently or through connections with previously engaged countries?
   \item How collaborative is mathematical oncology in comparison to other fields? Has this changed over time?
   \item How has mathematical oncology adapted to wider research breakthroughs? 
   Has the research focus of mathematical oncology changed over time?
   \vspace{-.1cm}
\end{enumerate}
\vspace{-.1cm}

\subsection{Two-stage bibliometric methodology}
\vspace{-.1cm}
To study research questions R1 to R4, and the development of mathematical oncology within its disciplinary context as a subfield of mathematical biology, we adopt a journal-first approach that follows a two-stage bibliometric methodology \citep{Waltman2013}. 
In the first stage, we perform a data-driven selection of focus journals and, in the second stage, we classify articles from the these as belonging to mathematical oncology or not. 

While a direct keyword-based article selection method (that bypasses the journal filter) would yield a larger dataset by capturing all articles that mention some specified cancer and mathematics keywords, it would also introduce significant keyword bias and noise. 
Such unfiltered methods often include articles that reference keywords superficially, without necessarily engaging in the epistemic practices of the field studied under the bibliometric lens \citep{Chen2016}. 
Via our two-stage process, we are instead able to focus on core mathematical biology journals, ensuring that the articles analysed are, in fact, grounded in the field of mathematical biology. 
This approach also provides a built-in reference set in the form of non-oncology articles from the same journals, allowing for direct comparison between mathematical oncology and its parent field mathematical biology after ensuring sufficient data -- articles -- in both groups \citep{Rogers2020}.

It is important to note that our journal-first approach excludes all mathematical oncology work published outside our focus journals, which represents a substantial portion of the field. 
As the field has increasingly integrated with practical, medical, and industrial applications, mathematical oncology articles are now more frequently published in multidisciplinary journals, such as Proceedings of the National Academy of Sciences (PNAS), life science journals like Nature Communications, and specialised cancer journals such as Cancer Research.

\vspace{-.1cm}
\newpage
\subsection{Journal selection}
\vspace{-.1cm}
\label{sec:journal_sel}
In our journal-first approach, our first task is to identify mathematical biology journals that frequently publish mathematical oncology articles. 
To identify these journals, we search the Web of Science Core Collection (WoSCC) for articles containing specific keywords: both a cancer-related term and a term containing the substring \texttt{mathematic*} (Figure \ref{fig:flowchart_article_selection}a). 
We use the WoSCC as our main database as it has excellent coverage of the mathematical biology literature (Supplementary Material, S3).

For all journals that have published more than 100 articles that match the keywords, we evaluate their relevance to mathematical biology in a twofold manner. 
First by identifying their journal classification according to the Quacquarelli Symonds (QS) scheme \citeyearpar{QSranking}, and second by examining their self-authored aims and scopes as written on the journal websites (Supplementary Material, S1). 
Our data-driven assessment reveals five mathematical biology journals with more than 100 mathematical oncology articles. 
We will refer to these as our {\it focus journals} throughout this study, and they are: the Bulletin of Mathematical Biology (BMB), the Journal of Mathematical Biology (JMB), the Journal of Theoretical Biology (JTB), Mathematical Biosciences (MB), and Mathematical Biosciences and Engineering (MBE), as listed in Table \ref{tab:math_bio_journals}. 
\vspace{.1cm}

\begin{table}[h]
\centering
{\small
%\resizebox{\textwidth}{!}{%
\begin{tabular}{|>{\raggedright\arraybackslash}p{7cm}|>{\raggedright\arraybackslash}p{8cm}|}
\hline
\textbf{\shortstack[l]{Journal name\\(ISO4)}}& \textbf{\shortstack[l]{Established\\(first mathematical oncology article)}}   % & \textbf{Publisher} 
\\ \hline
\shortstack[l]{Bulletin of Mathematical Biology\\(Bull. Math. Biol.)}& 1972 \citep{swan1976solution} %& Springer 
 \\ \hline
\shortstack[l]{Journal of Mathematical Biology \\(J. Math. Biol.)} & 1974 \citep{merrill1984stochastic} %& Springer 
 \\ \hline
\shortstack[l]{Journal of Theoretical Biology\\ (J. Theor. Biol.)} & 1961 \citep{gause1961biological} %& Elsevier 
 \\ \hline
\shortstack[l]{Mathematical Biosciences \\(Math. Biosci) }& 1967 \citep{bellman1967quasilinearization} %& Elsevier 
 \\ \hline
\shortstack[l]{Mathematical Biosciences and Engineering\\ (Math. Biosci. Eng.)} & 2004  \citep{banks2004modeling} %& American Institute of Mathematical Sciences Press 
 \\ \hline
\end{tabular}
%}%end small
\caption{\textbf{The mathematical biology focus journals researched in this study.} 
The Bulletin of Mathematical Biology was originally established in 1939 under the name Bulletin of Mathematical Biophysics.}
\label{tab:math_bio_journals}
}
\end{table}

\vspace{-.2cm}

\subsection{Data collection}
We collect article metadata from the WoSCC using a WoS API \citep{clarivate_wos_api}.  
Specifically, we download metadata for all articles published in our focus journals up to and including the year 2024. 
For each article, we obtain the title, publication year, author names and affiliations, abstract, author keywords, references, and a list of items that cite the article. 
Articles published in MB between 1967 and 1975 are not indexed in the WoSCC, thus we download their metadata from Scopus.

\subsection{Data cleaning and article classification}
\label{sec:article_selection}
\begin{figure}[h!]
    \vspace{-.2cm}
    \centering
    \includegraphics[width=\linewidth]{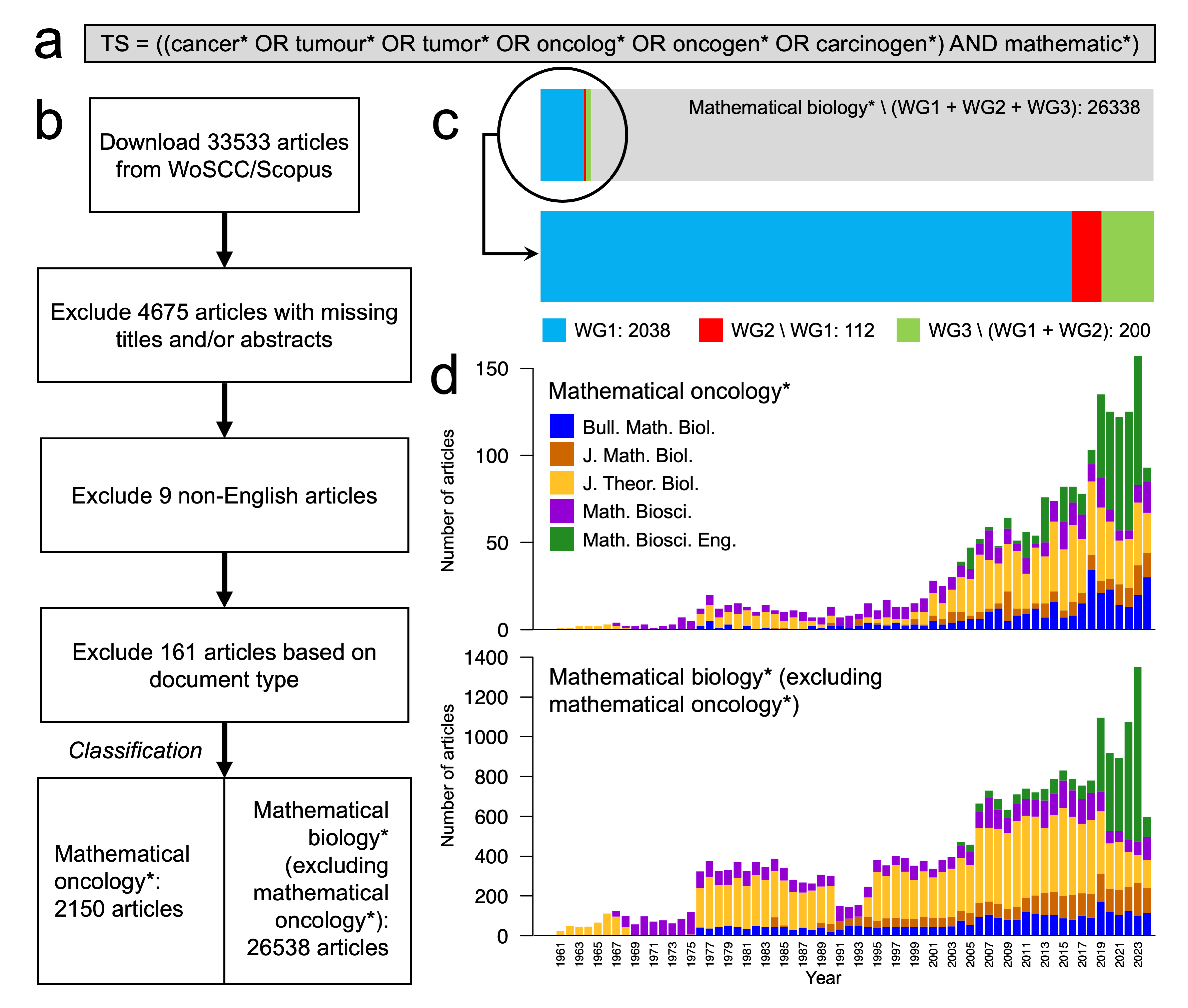}
    \vspace{-.8cm}
    \caption{{\bf Article selection and classification.}
    (a) The topic search (TS) query used in the WoSCC to identify mathematical oncology articles and inform the journal selection.
    (b) The data collection, cleaning and classification procedure.
    (c) The proportion of candidate keyword groups (WG) 1-3 proposed in the article classification, with article counts given after the colons. 
    (d) Annual article counts for the focus journals following the article classification in (b).
    }
        \vspace{-0.4cm}
    \label{fig:flowchart_article_selection}
\end{figure}

Before proceeding with our bibliometric analysis, we clean the downloaded metadata in three steps:
first, we remove articles with missing titles and/or abstracts;
next, we omit non-English articles; 
and finally, we exclude documents classified as editorial material, notes, corrections, letters, biographical-items, and reprints \ref{fig:flowchart_article_selection}b. 
As a result, we end up with only original research and review papers. 

After data cleaning, we classify each article as belonging to one of two datasets: \textit{mathematical oncology*} or \textit{mathematical biology* (excluding mathematical oncology*)}. 
Throughout this article, we will use asterisks to denote these datasets (as opposed to the full research fields). 
To classify which of the two datasets an article belongs to, we use keyword matches in the article's  title, abstract, and author keywords.
If the article contains at least one cancer-class-keyword match, it belongs to the mathematical oncology* dataset and, otherwise, it does not.

It is not trivial to determine which keywords to include in bibliometric article classifications, as too narrow keywords might exclude articles, and too broad keywords might over-include articles in some specified class \citep{Chen2016}. 
To approach this problem for our specific dataset, we formulate three candidate keyword groups (WGs) where
WG1 includes the general cancer terms listed in Figure \ref{fig:flowchart_article_selection}a;  
WG2 includes a list of 57 cancer types such as \texttt{leukemia}, \texttt{melanoma}, and \texttt{glioma}; and   
WG3 includes a list of 19 words that are commonly mentioned alongside cancer such as \texttt{chemotherapy} and \texttt{radiotherapy}. 
A full list of keywords for all WGs is available in the Supplementary Material (S4). 
The frequency of articles that contain keywords in the studied WGs is shown in Figure \ref{fig:flowchart_article_selection}c, and the resulting annual dataset sizes, after data cleaning and the final article classification, are shown in Figure \ref{fig:flowchart_article_selection}d.

To determine which WGs to include in our classification, we evaluate their predictive performance manually by reading a subset of randomly selected article abstracts from each WG and, when needed, the full manuscripts. 
For WG1, WG2$\backslash$WG1, and WG3$\backslash$(WG1$\cup$WG2), we examine 100 articles predicted to be in mathematical oncology*, and 100 articles predicted to be in mathematical biology* (excluding mathematical oncology*). 
In brief, we found that using WG1$\cup$WG2  yielded the best performance -- neither over-excluding nor over-including articles in the mathematical oncology* dataset. 
Accordingly, we classify any article containing at least one keyword from WG1$\cup$WG2 in the title and/or abstract as a mathematical oncology* article.

\vspace{.4cm}

\vspace{.4cm}

\newpage
Leading up to this chosen classification, our evaluation showed that using WG1 as a classifier achieves 99\% precision, i.e., 99\% of the articles predicted as mathematical oncology* were also classified as such by human reviewers, but misses multiple relevant articles from WG2$\backslash$WG1.
In contrast, WG3$\backslash$(WG1$\cup$WG2) achieves a 96\% false discovery rate, i.e., 96\% of articles predicted as mathematical oncology* were not classified as such by humans, and is consequently excluded from the classification keywords. Detailed 
quantitative results from the keyword classification evaluations are presented in a confusion matrix in the Supplementary Material (S4).
\vspace{.15cm}

\subsection{Journal classification for citation analysis}
\vspace{.1cm}
\label{sec:journal_cat}
To analyse citation flows, we categorise citing journals into seven disciplinary areas of interest. 
Beyond our focus journals, these are: mathematics and life sciences, mathematics, life sciences, 
multidisciplinary, STEM and life sciences, and other. 
To enable this categorisation to be fine-grained in subject areas close to mathematical oncology, and coarse-grained elsewhere, we customise a journal classification that builds on the well-established 
QS subject rankings \citeyearpar{QSranking} and the All Science Journal Classification (ASJC) scheme \citep{elsevierASJC}. 
In the ASJC scheme, a journal is assigned one or more category codes based on journal metadata, and the 
QS subject rankings further aggregate these codes into subject and broader faculty areas. 
Here, we use ASJC-QS classifications to directly map each citing journal to one of our seven disciplinary categories (Supplementary Material, S2).

\vspace{.15cm}
\subsection{Data visualisation}
We use R \citep{R} and the R-package biblioshiny \citep{aria2017bibliometrix} to visualise bibliometric data. 
Instructions on how to generate the figures presented in this article are available on the study's GitHub repository. 
\vspace{.15cm}

\section{Results}

\label{sec:results}

\subsection{On interdisciplinarity in mathematical oncology}
\vspace{-.1cm}
\label{sec:intedis}
As a first research question (R1) we ask: {\it Which disciplines influence and are influenced by mathematical oncology?} 
Equivalently put, is mathematical oncology a research field for (a) mathematicians, (b) cancer researchers, (c) mathematical biologists, (d) others, or (e) all of the above? 
Here, we take a discipline-centered bibliometric approach to answer this question. 

\begin{figure}[h!]
     \centering
     \includegraphics[width=1\textwidth]{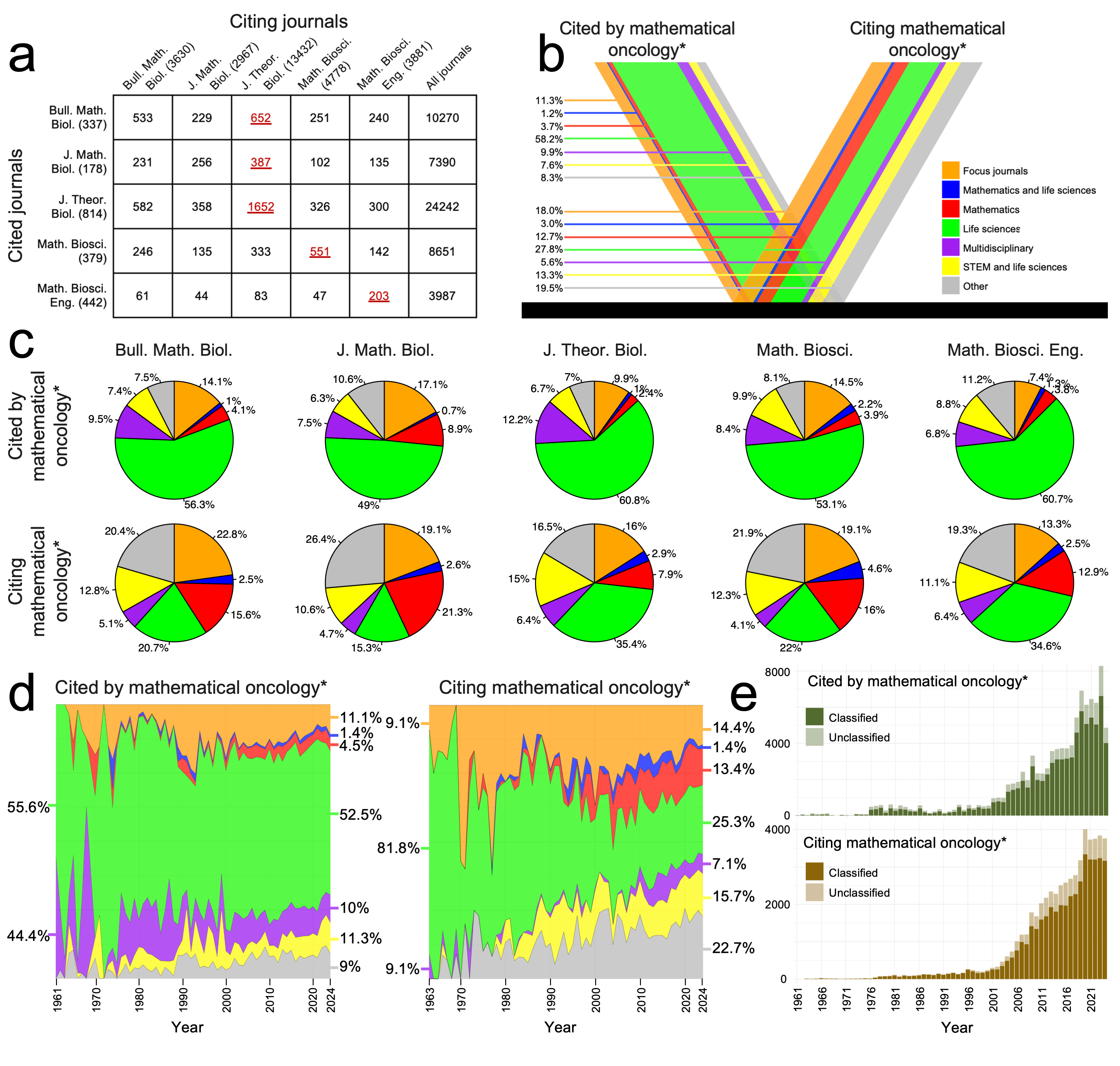}
     \caption{
     \small
     {\bf Interdisciplinarity in mathematical oncology.} 
     (a) The citation matrix shows how often mathematical oncology* articles (rows) are cited by mathematical biology* articles (columns). The right-most column shows the number of row-wise citations from all journals (including those beyond our focus journals). The total number of mathematical oncology* articles per journal are shown in row-legend parentheses, and the total number of mathematical biology* articles per journal are shown in column-legend parentheses. 
     %
     %Matruix element divided by math bio* parenthesis
     %
     The maximum number of citations per row are red and underlined. 
     (b) The citation-V shows the percentage of articles in different disciplinary categories that are cited by mathematical oncology* (left), and that cite mathematical oncology* (right) for all focus journals between 1961 and 2024. 
     (c) The results in (b) are shown for one focus journal at a time in pie charts. 
     (d) The results in (b) are shown over time. 
     For each citation, it is the time (year) that the citation was made that is shown in one of the plots.  
     (e) Annual counts for articles that are cited by (top) and that cite (bottom) mathematical oncology*. Counts for both discipline-classified and unclassified journals are shown.  
     {\it The notation mathematical oncology* refers to the studied dataset.}
    }
    \label{fig:interdisciplinarity}
 \end{figure}

As a preparatory step, we map out inter-citation patterns between our five mathematical biology focus journals. 
The citation matrix in Figure \ref{fig:interdisciplinarity}a demonstrates that mathematical oncology articles published in JTB, MB, and MBE have the most citations from themselves, whereas mathematical oncology articles in BMB and JMB -- the flagship journals of the Society for Mathematical Biology and the European Society for Mathematical and Theoretical Biology, respectively -- obtain most of their citations from JTB. 
This demonstrates the societies’ intradisciplinary reach, although it is partly explained by JTB being the focus journal with the largest number of articles. 
Indeed, the effect of journal size on self-citations has been observed in scientometric studies \citep{tacskin2021self}. To mitigate this effect, we include an alternative version of the citation matrix in the Supplementary Material (S5), using normalisation by the number of mathematical biology* articles in each respective journal. 
The normalised matrix shows that all focus journals, except JTB, account for the highest fraction of citations within themselves, indicating strong internal citation patterns even after adjusting for journal size in the mathematical biology* dataset. 

\newpage

\vspace{-0.05cm}

Next, to identify the disciplines that {\it influence} and {\it are influenced by} mathematical oncology, we extend our scope to study all articles that {\it are cited by} and/or {\it cite} articles from our mathematical oncology* dataset. 
We classify these articles into disciplines based on their journals, following the scheme outlined in Section \ref{sec:journal_cat}. 
For each discipline, journals that most frequently are cited by and cite the mathematical oncology* dataset are listed in the Supplementary Material (S5).
With our scheme, we are able to classify 81\% of the articles that are cited by mathematical oncology*, and 82\% that cite mathematical oncology*. 

\newpage
\vspace{.5cm}
The spectra of disciplines that are cited by, and cite, the full mathematical oncology* dataset are shown in Figure \ref{fig:interdisciplinarity}b through the citation-V. 
The figure reflects that the vast majority of articles cited by mathematical oncology* belong to life science journals (58.2\%). Thereafter, the focus journals (11.3\%) and multidisciplinary journals (9.9\%) are the most cited. 
Notably, only 3.7\% of the cited articles are published in mathematics journals. 
When we instead examine which articles cite mathematical oncology*, the disciplinary composition is more diffused, meaning citations come from a broader and more evenly distributed range of fields. 
Most strikingly, the percentage of articles from life science journals has been reduced to 27.8\%, while articles from mathematics journals have increased to 12.7\%. 
In between, articles from the focus journals account for 18.0\% of the citations. 
Thus, to paraphrase Reed's article {\it Mathematical biology is good for mathematics} \citeyearpar{Reed2015}, our quantitative results suggest that mathematical oncology is good for, or at least cited by, mathematics.

When breaking this result down to a journal level, we see that citing trends differ among our focus journals (Figure \ref{fig:interdisciplinarity}c). 
While life science articles account for the majority of citations in all focus journals, ranging from 49.0\% in JMB to 60.8\% in JTB, mathematics articles account for 8.9\% of in JMB, but less than half of that in all other journals. 
Even on a journal level, the diffusive effect of mathematical oncology is clear as the composition of disciplines that cite mathematical oncology* articles is significantly more evenly distributed than the composition of disciplines cited by mathematical oncology*. 

The last result raises the question: {\it is this diffusive effect something that has emerged over time?}
To approach this research question, we plot the proportion of articles that are cited by, and cite, mathematical oncology* over time in Figure \ref{fig:interdisciplinarity}d. 
The plot demonstrates a clear shift over time; 
while life science disciplines dominated both the ``cited by'' and ``citing'' categories in the 1960s, the representation of mathematics has, overall, grown over time and has had a steady presence since the early 2000s. 
The representation of focus journals in both categories has remained largely consistent since the 1990s, as have the related disciplinary categories ``mathematics and life sciences'' and ``STEM and life sciences''. 
These results interestingly contradict the theory that research fields tend to become increasingly internalised and interact less with external disciplines as they mature \citep{singh2022quantifying}, as is further elaborated on in the Discussion (Section \ref{sec:discussion}). 
To allow for relating the percentage-based results reported in this section to absolute numbers, we plot the total number of mathematical oncology* references and citations in Figure \ref{fig:interdisciplinarity}e. 
The corresponding results for each individual focus journal are available in the Supplementary Material (S5).

%\newpage

\vspace{0.5cm}

%%%%%%%%%%%%%%%%%%%%%%%%%%%%%%%%%%%%%%%%%%%%%%%%%%%%%%%%%%%%%%%%%%%%
        
\subsection{On internationality in mathematical oncology}
\label{sec:internationality}
\begin{figure}[h!]
 \vspace{1cm}
     \centering
     \includegraphics[width=1\textwidth]{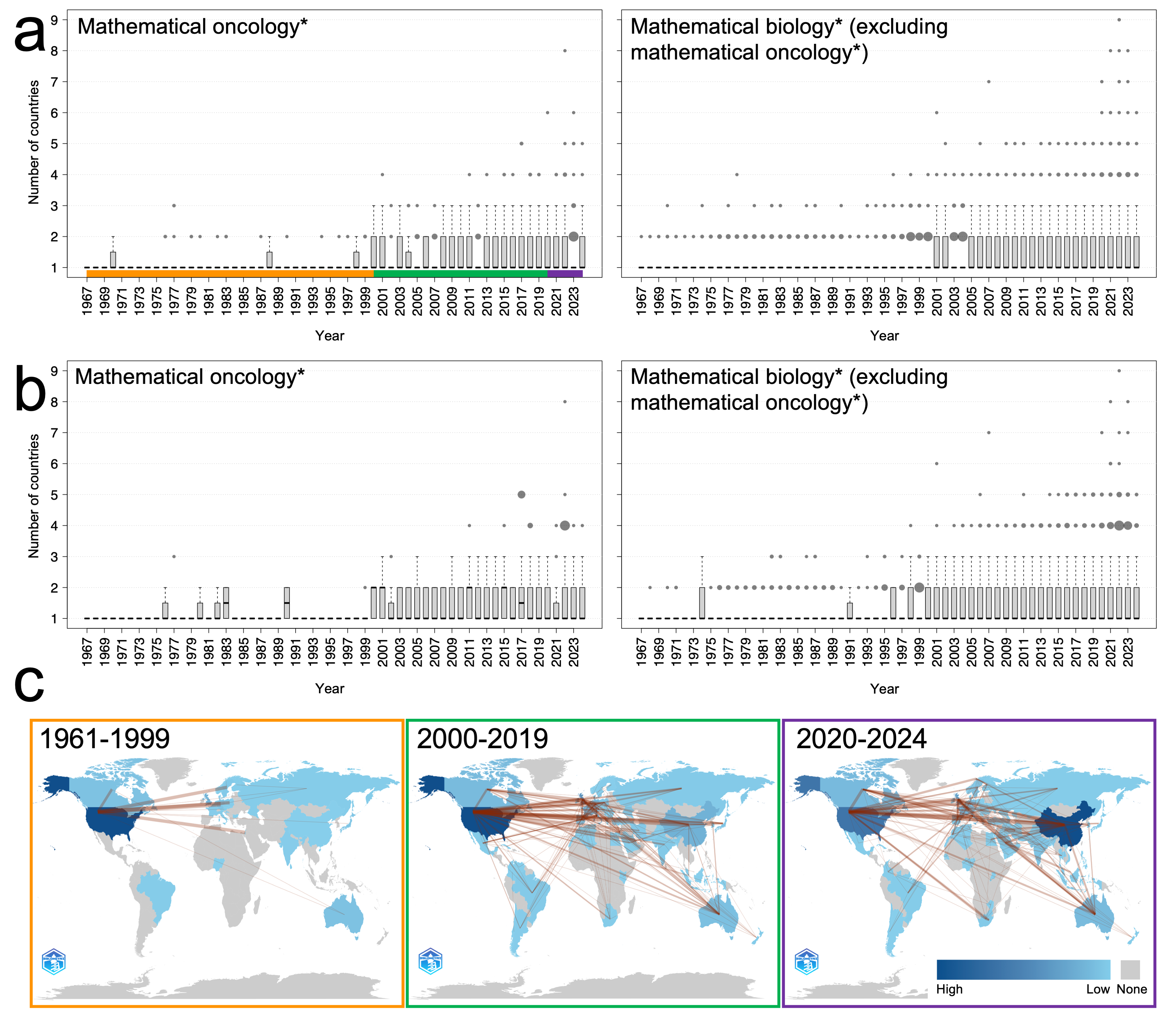}
     \caption{{\bf Internationality in mathematical oncology.}
     The box plots show the number of unique author affiliation-countries per article over time 
     for (a) all articles in the datasets, and (b) articles with above-average citations published that year.
       In the box plots, black lines indicate the medians; boxes show the interquartile ranges (IQR); whiskers extend to the smallest and largest data points within 1.5×IQR from the box edges; and black dots represent outliers, with dot size proportional to their frequency.
    (c) The heatmaps show country-wise contributions (number of articles) to the mathematical oncology* dataset for three different time eras. 
    Red lines demonstrate pairwise collaborations between countries, where the line thickness is proportional to collaboration frequency.
    Heatmap colours and line thicknesses are normalised for each time period. 
    {\it The notations mathematical oncology* and biology* refer to the studied datasets.}
     }
     \label{fig:internationality}
 \end{figure}

\vspace{0.3cm}
Motivated by research question R2, we next set out to study how international mathematical oncology is by analysing author affiliation countries in article metadata.
The box plots in Figure \ref{fig:internationality} show how many affiliation countries are represented per mathematical oncology* article, and how these numbers have changed over time since 1967, the first year with available affiliation data. 
To situate our results in a broader context, we also generate the corresponding box plots for the mathematical biology* (excluding mathematical oncology*) dataset. 
For both datasets, we further extract the articles with above-average (per publication year) citations to investigate whether there is any link between an article's country representation and its citation count. 
We do this by comparing medians (Q2) and the lower and upper quartiles (Q1 and Q3) across datasets. Note that these quartile values take half-integer values in the event of ties.

\newpage

We find that the mathematical oncology* dataset demonstrates a {slight} correlation between country counts and citation counts: in {7} out of 58 {years}, articles with above-average citation counts have a higher median number of affiliation countries than the all articles, and in 11 of 58 years they have a higher Q3 value. 
The reverse is true in only 0 and 4 years, respectively. We further observe a similar correlation in the mathematical biology* (excluding mathematical oncology*) dataset: in 7 out of 58 years, articles with above-average citation counts have a higher Q3 value than all articles combined in the dataset, with the reverse never being true. 

\begin{samepage}
While the Q1 and Q2 values have remained 1 over time, for all articles in both datasets, Figure \ref{fig:internationality}a demonstrates a slow increase in country count over time. 
This is reflected by the strong Pearson correlation coefficients for Q3 between 1961 and 2024 (0.7162 and 0.8321 for the mathematical oncology* and mathematical biology* (excluding mathematical oncology*) datasets respectively)  \citep{mukaka2012guide}.
Moreover, the most notable shifts in country affiliation counts are observed around the years 2000 and 2020, and we therefore
split the studied time period into three eras: the pre-millennia era (1961-1999), the early 21st century (2000-2019), and the 2020s (2020-2024). 
We speculate that the last split is related to the COVID-19 pandemic which notably impacted the culture of STEM research \citep{Heo2022}. 
Pearson correlation coefficient values for Q1, Q2, and Q3, across all sub-datasets and time eras, are provided in the Supplementary Material (S6). 

\end{samepage}

For each era, countries that contribute to the mathematical oncology* dataset are highlighted on world maps (Figure \ref{fig:internationality}c), with colours representing the number of article contributions per country.
The maps are generated in Bibliometrix \citep{aria2017bibliometrix}. For each article, every unique author affiliation contributes one count to the country heatmap, and each distinct country pair contributes one count to the collaboration edges. For example, if an article has two authors from different institutions within the same country, that country contributes two counts. Conversely, if an article includes at least one author from country A and one from country B, it is counted as a single A-and-B collaboration, regardless of the number of affiliations within each country.
Both country-wise article counts and pairwise collaboration frequencies are available in tabulated form in the Supplementary Material (S6). 

Together, the three maps indicate a geographical decentralisation of mathematical oncology* over time, following general scientific globalisation trends \citep{Dong2017}. 
It is interesting to note that when new countries appear on the map in the last two eras, they typically do so as connected nodes to previously engaged countries.  
This pattern can, for instance, be observed across three continents in the early 21st century map, by panning from left to right and noting the addition of Argentina, South Africa, and New Zealand. 
Although these results demonstrate correlation, not causation, they still signify the importance of enabling research connections to support globalisation.  
This is in line with previous studies on sustainable partnerships between the Global North and South for research \citep{Kunert2020} and education \citep{Mahdjoub2023} purposes.
As such, organisations aiming to support the globalisation of mathematical oncology can use the findings reported in this section in two important ways: 
first, to identify underrepresented countries; and second, to motivate financial and infrastructural support for international collaboration networks.
%
%%%%%%%%%%%%%%%%%%%%%%%%%%%%%%%%%%%%%%%%%%%%%%%%%%%%%%%%%%%%%%%%%%%%

%\vspace{.5cm}
\subsection{On collaboration in mathematical oncology}

\begin{figure}[!ht]
     \centering
     %\vspace{.5cm}
     \includegraphics[width=1\textwidth]{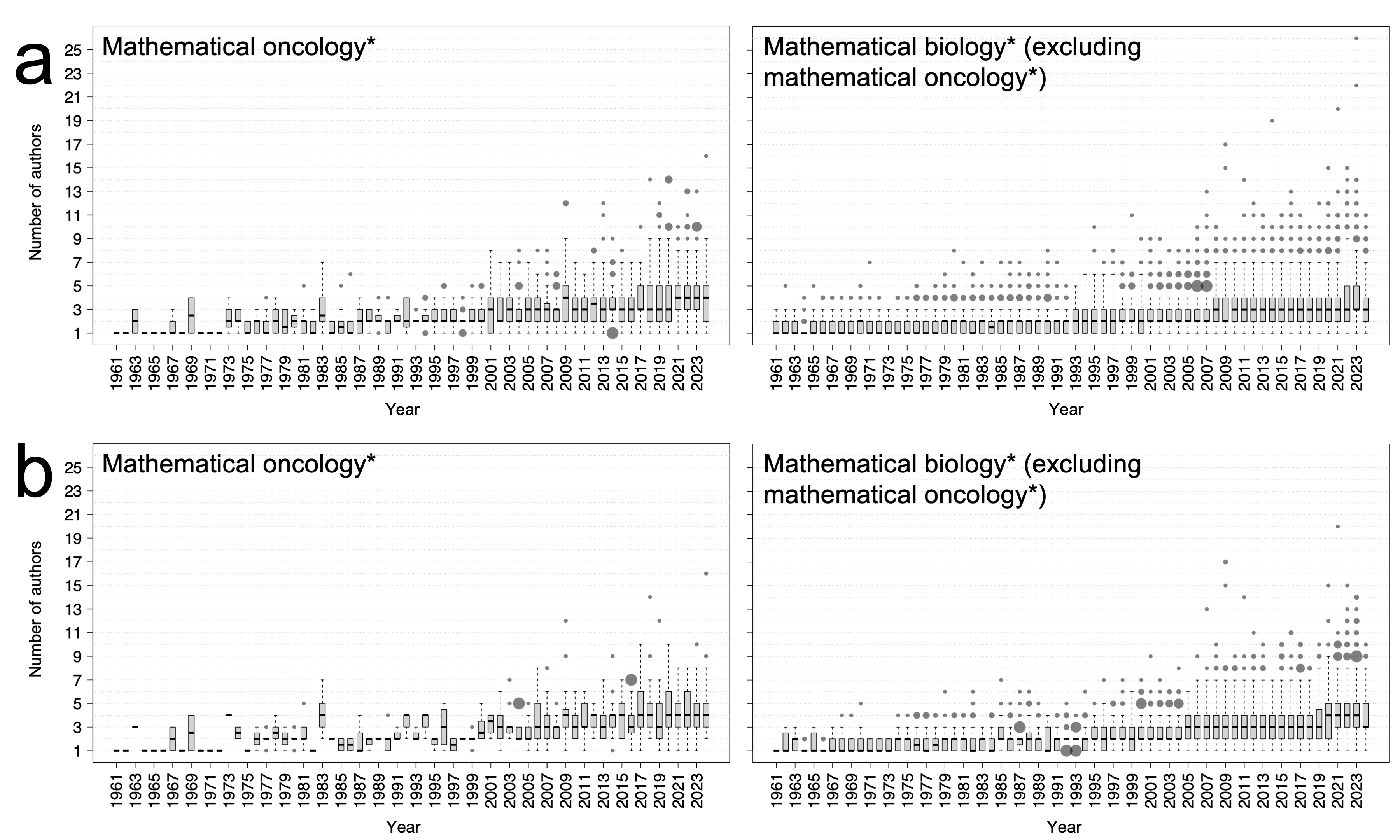}
     \caption{ {\bf Collaboration in mathematical oncology.} 
     The box plots show the number of authors per article over time 
     for (a) all articles in the datasets and (b) articles with above-average citations published that year.
     In the box plots, black lines indicate the medians; boxes show the interquartile ranges (IQR); whiskers extend to the smallest and largest data points within 1.5×IQR from the box edges; and black dots represent outliers, with sizes proportional to their frequency.
    {\it The notations mathematical oncology* and biology*, refer to the studied datasets.}
    }
     \label{fig:collaboration}
 \end{figure}

   %  \vspace{.5cm}

Continuing the discussion on collaborations and turning to research question R3, we now analyse team sizes in mathematical oncology* studies, again building on author metadata. 
This is particularly interesting as the field overlaps both mathematics, a discipline historically dominated by small teams and single-author articles, and biomedicine, which typically involves larger collaborative teams \citep{porter2009science}. 

Taking a similar approach as in the previous section, we use box plots to visualise how the number of authors per article has changed over time in Figure \ref{fig:collaboration}a. 
The figure reveals a slow and steady increase over time in the number of authors per article in both the mathematical oncology* and mathematical biology* (excluding mathematical oncology*) datasets, as is quantified by strong Pearson correlation coefficient values over 0.8 for Q1, Q2, and Q3 during 1961-2024 for both datasets \citep{mukaka2012guide}. 
These results follow general trends for life science publications \citep{Rao2022} and science more broadly \citep{lariviere2015long, fortunato2018science}. 
Comparing the two plots in Figure \ref{fig:collaboration}a we also see that, since 2017, mathematical oncology* articles have generally had slightly larger team sizes (in terms of interquartile ranges) than mathematical biology* articles as a whole. 
We conjecture that this may be related to the broader trend that, within the biomedical literature, publications directly or closely related to clinical implementation have experienced the highest increases in team sizes since the turn of the millennium \citep{Jakab2024}.

We next focus on citation counts again and see that, for the mathematical oncology* articles, the median number of authors is higher in articles with above-average citation counts compared to all articles in 22 out of 64 years, and lower in only 3. 
A similar pattern appears for the lower and upper quartiles, with Q1 and Q3 values being higher for above-average cited articles in 21 and 16 years, respectively, and lower in only 4 and 13 years. 
These results signify that articles with above-average citation counts are generally associated with larger sized teams.

In a broader scientific context, bigger team sizes are generally correlated with higher impact through citations, an outcome that is not merely a consequence of self-citations, but may also be a product of the pooled knowledge that goes into a study when researchers work together, especially across disciplines \citep{Larivire2015}. 
In mathematical oncology studies, where it is not uncommon for mathematicians, data scientists, biologists, and clinicians to collaborate, it is thus natural to reason that increased team sizes would come with a citation-count benefit. 
A similar, but slightly weaker, correlation between team sizes and citation counts is identified for the mathematical biology* (excluding mathematical oncology*) articles. This is shown in Figure \ref{fig:collaboration}, 
where the median number of authors is higher for articles with above-average citations compared to all articles in the dataset in 13 out of 64 years, with the reverse being true in only 3 years. Similarly, Q1 and Q3 are higher for above-average cited articles in 10 and 12 years, respectively, and lower in only 1 and 4 years.
All Pearson correlation coefficient values for Q1, Q2, and Q3, across each dataset and time era, are provided in the Supplementary Material (S7).

\begin{figure}[h!]
     \centering
     %\vspace{2cm}
    \includegraphics[width=1\textwidth]{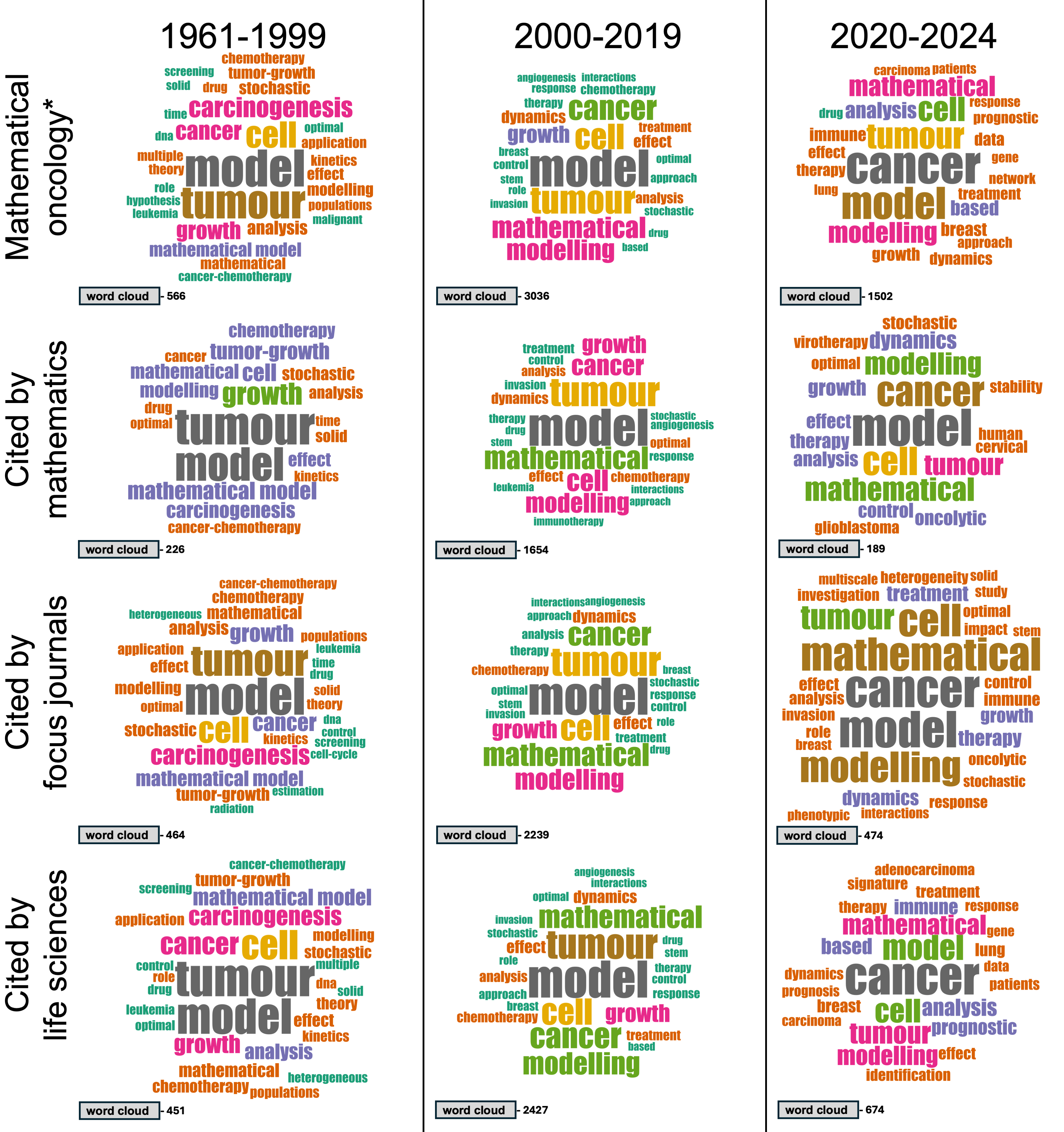}
    \vspace{-0.1cm}
       \caption{{\bf Trending topics in mathematical oncology analysed through title word frequency.}
     In each subplot, the bar chart shows the frequency of the collective word cloud terms.
     The word clouds contain the 25 most frequent terms with sizes proportional to their frequency. 
     Data for all mathematical oncology* abstracts, per time period, are shown in the top row. 
     Data for articles with at least one citation in mathematics, focus, and life science journals are shown in the three rows below, respectively. 
     }
     \label{fig:word_clouds_title}
 \end{figure}

\begin{figure}[h!]
     \centering
     %\vspace{8cm}
    \includegraphics[width=1\textwidth]{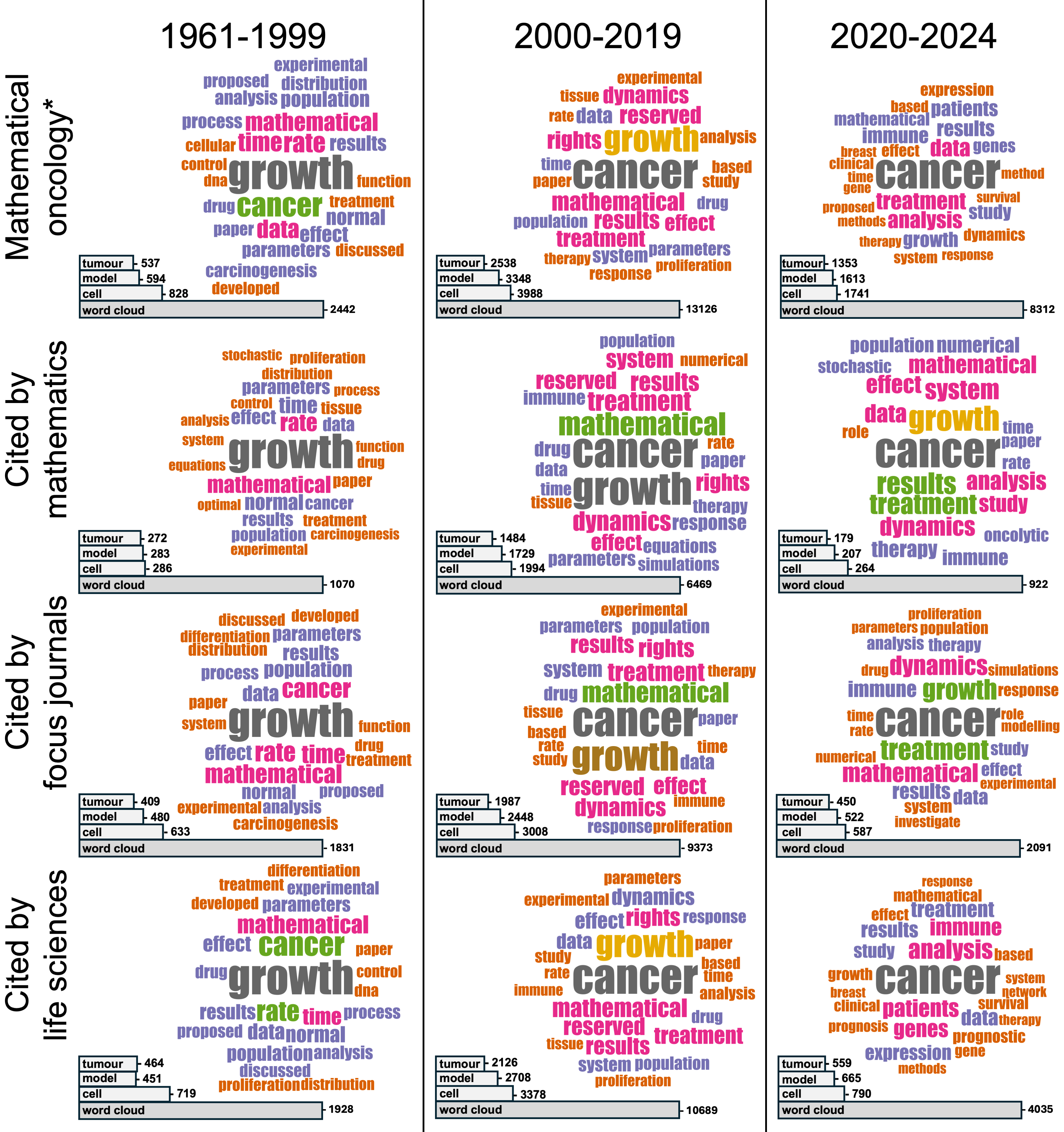}
     \caption{{\bf Trending topics in mathematical oncology analysed through abstract word frequency.}
     In each subplot, the bar charts show the frequency of the terms \texttt{tumour}, \texttt{model}, \texttt{cell}, and the collective word cloud terms. 
     The word clouds contain the 25 most frequent terms after similarity pruning (removing \texttt{tumour}, \texttt{model}, \texttt{cell}) with sizes proportional to their frequency. 
     Data for all mathematical oncology* abstracts, per time period, are shown in the top row. 
     Data for articles with at least one citation in mathematics, focus, and life science journals are shown in the three rows below, respectively. 
     }
     \label{fig:word_clouds_abstract}
 \end{figure}

%%%%%%%%%%%%%%%%%%%%%%%%%%%%%%%%%%%%%%%%%%%%%%%%%%%%%%%%%%%%%%%%%%%%
\newpage
\begin{samepage}

\vspace*{-.3cm}
\subsection{On trending topics in mathematical oncology}
%\vspace{-.35cm}
\vspace{0.3cm}
We now set out to investigate research question R4 by identifying trending research topics in mathematical oncology and how these have varied over time.  
We do this via word clouds, which can be used to flexibly visualise term frequency in bibliometric datasets without predefined categories or keywords.  
The word clouds in the top rows of Figures \ref{fig:word_clouds_title} and \ref{fig:word_clouds_abstract},
respectively, show the frequency of words in the titles and abstracts of our mathematical oncology* dataset over the three time periods defined in Section \ref{sec:internationality}. 
The word cloud data are generated in Bibliometrix, which automatically filters out common stopwords (e.g., ``and", ``the", ``or") from the analysis. 
Additionally, to reduce redundant counting of terms, we created our own list of synonyms which is provided in the Supplementary Material (S8).
Tabulated word cloud data, and corresponding histograms, are also available in the Supplementary Material (S8).

Extending our prior investigation into which disciplines consume (are influenced by) mathematical oncology (Section \ref{sec:intedis}), Figures \ref{fig:word_clouds_title} and \ref{fig:word_clouds_abstract} contain word clouds for all articles cited by (at least one article in) in three journal categories: our focus journals, mathematics, and life science journals. 
Together, these word clouds highlight similarities and differences between time periods and consumer groups. 
We first note that the words \texttt{cell}, \texttt{model}, and \texttt{tumour} have consistently been in the top three abstract words (and the top six title words) for all time periods and consumer groups, indicating stability of the field. 
To prevent these high-frequency words from dominating the abstract word clouds and obfuscating information (Figure \ref{fig:word_clouds_abstract}), we display them through bars instead of words. 

Moreover, the word clouds reveal shifts in trending topics that are homogeneous across consumer groups; \texttt{angiogenisis} is present in all title clouds in the early 2000s but not elsewhere, and \texttt{immune} appears in all consumer-group abstract clouds for the last two time eras.
We conjecture that these observations are related: the latter reflects the series of immunotherapy breakthroughs that have transformed cancer treatments since the 2000s \citep{Dobosz2019}, and the former that anti-angiogenic agents were first FDA approved in 2004 and are now predominantly used and studied in combination with other therapies, such as immunotherapy \citep{Ansari2022}. 
Another word that is present in all word clouds for the abstracts is \texttt{data}, but only in the 2020s does the word feature in the title clouds -- and then only for the complete and life science consumer groups. This suggests that, although data have always been an integral part of the field, it is only recently that authors have promoted (or been awarded for promoting) the use of data in article titles, which is presumably linked to the fact that cancer research is becoming increasingly data-driven \citep{Jiang2022}. 

The finding that life science, but not mathematics, journals cite articles with \texttt{data} in the title indicates a split in interests between the two consumer groups. 
Meanwhile, the word \texttt{equations} is present only in the cited-by-mathematics abstract word clouds. Similarly, the related words \texttt{numerical} and \texttt{simulation} only appear in the word clouds for the mathematics and mathematical biology (focus journal) groups. 
In line with this observed split, Fawcett and Higginson \citeyearpar{fawcett2012heavy} argue that heavy use of equations impedes communication in biology articles, whereas Chitnis and Smith  \citeyearpar{chitnis2012mathematical} rebut this statement by claiming that it is mathematical illiteracy that is hindering the use of equations in biology.
Regardless of which stance one takes in this debate, our analysis reveals distinct differences between articles cited by mathematics versus life science journals, with those cited by mathematical biology lying somewhere in between. 
%

%\begin{samepage}
Further, the title word clouds for the complete dataset and the cited-by-life sciences suggest an increased specificity towards cancer types over time, as words like \texttt{breast} and \texttt{lung} increase in frequency.  
We also infer from the title data that articles cited by life science journals have recently become more focused on clinical applications, as the words \texttt{clinical}, \texttt{patients}, and \texttt{prognostic} first appear in the 2020s, where the last word also indicates a move towards using mathematical oncology for predictions
-- a task where machine learning models commonly outperform mechanistic ones, given sufficient data \citep{baker2018mechanistic}. 
Related trends towards data, machine learning and the encompassing field of AI have been observed in other areas of applied mathematics \citep{li2025mapping} and science overall \citep{Hajkowicz2023} through bibliometric studies. 

\end{samepage}
%\end{samepage}

%\vspace{0.7cm}

%\newpage
\section{Discussion}
\label{sec:discussion}
\vspace{-.2cm}
Our bibliometric research reveals that, over the past seven decades, mathematical oncology has undergone a significant evolution marked by increased interactions with other disciplines, geographical expansion, larger research teams, and greater diversity in research topics.
To further characterise the role that mathematical oncology plays in the larger knowledge system, we here draw on Kwon et al.'s framework \citeyearpar{kwon2017measure} which classifies citation flows and, by extension, interdisciplinary knowledge flows, into three types: aggregating, bridging, and diffusing. 
Aggregating articles cite two or more distinct source disciplines, whereas bridging articles cite one discipline and are cited by another, and diffusing articles are cited by multiple disciplines. 
From our data analysis, we find that mathematical oncology performs all three of these roles, but not symmetrically.  
It does not typically aggregate mathematics and oncology directly; instead it primarily draws on knowledge from the life sciences and adjacent disciplines that combine the life sciences with other areas of STEM, 
often integrating quantitative approaches into biological contexts. 
Furthermore, our observed diffusive and bridging citation patterns support Reed’s hypothesis that mathematical biology (and its subfields) benefits both mathematics and biology \citep{Reed2015}. 

%
%\newpage

These knowledge flow patterns contrast with those described by Singh et al. \citeyearpar{singh2022quantifying}, who find that scientific fields typically become more internally focused as they mature, citing external disciplines less frequently. 
Their findings align with Eddy’s \citeyearpar{eddy2005antedisciplinary} notion of `antedisciplinary' science. Using molecular biology as an example, Eddy argues that early work in new disciplines emerges outside existing disciplines and later consolidates into its own self-contained identity. 
Mathematical oncology, on the other hand, appears to mature without such disciplinary closure. Rather than narrowing its scope, the field has become more interdisciplinary over time; continually drawing on and contributing to both mathematics and the life sciences.
Mathematical oncology has, of course, developed a disciplinary identity of its own, as reflected by dedicated research communities, events, and education tracks, but the field continues to thrive through active engagement with both mathematics and the life sciences.

While citation flows illustrate disciplinary connections, they reveal little about {\it how} researchers engage with mathematical oncology. To this end, word clouds provide important complementary insight into the nature of mathematical oncology research that different disciplines interact with. 
As discussed in the previous section, mathematical oncology* articles published in the last five years that are cited by life science journals increasingly emphasise the use of data in their titles, and tend to focus on clinical applications and prediction. 
This may reflect the broader rise in data availability within cancer research \citep{Jiang2022}, but also signals a shift in emphasis; while data has long been part of abstracts in the field, its recent prominence in titles suggests changing expectations around what makes an article valuable (or citable) in the life sciences.

In contrast, articles cited by mathematics journals more often feature terms such as \texttt{equations}, \texttt{numerical}, and \texttt{simulation}. 
While this observation, as discussed in the previous section, indeed reflects the debate put forward by \citep{fawcett2012heavy} and \citep{chitnis2012mathematical} centered around the presentation of equations in theoretical biology papers, it also touches upon an even older discussion in mathematical biology on the values and trade-offs in modelling \citep{Levins1966}.
Levins argued that, when modelling biological systems, no model can simultaneously maximise generality, realism, and precision. Our word cloud analysis indicates that, over the last years, the life science community increasingly value models focusing on precision, whereas the mathematics community seem to focus on models that Levins would categorise as emphasising generality. 

These patterns may point to the beginning of a broader disciplinary divergence, where mathematical oncology is increasingly split between data-driven, clinically oriented work and more abstract, mathematically focused modelling. 
While we cannot determine an article's modelling goals from keyword analysis alone, the observed shifts in terminology raise important questions about which kinds of models different communities find useful or persuasive. 
What makes a mathematical model of a biological system {\it good} -- mechanistic insight, predictive accuracy, or clinical relevance? 
Or; are mathematical models that prioritise mechanistic understanding of biological processes becoming less relevant in a research landscape increasingly dominated by machine learning models focused on prediction? Baker et al. \citeyearpar{baker2018mechanistic} explore this tension and propose hybrid approaches that combine mechanistic modelling with machine learning techniques as a way forward. Complementing this discussion, Gyllingberg et al. \citeyearpar{gyllingberg2023lost} argue that for mathematical models to remain relevant to the life sciences, the field of mathematical biology must avoid an overemphasis on mathematical rigour and instead adopt a more pluralistic, biologically grounded modelling practice.

\newpage

Although our bibliometric research offers a window into how different communities engage with mathematical oncology, our results are based solely on citation patterns and keyword matches in article metadata.
In future work, further information on interdisciplinarity in mathematical oncology could be gathered from author affiliations and training backgrounds, in line with previous work by Abramo et al. \citeyearpar{Abramo2012}. 
To better segment our findings, future studies could also apply established interdisciplinarity metrics, such as the Rao–Stirling index \citep{stirling2007general}, to more precisely quantify how interdisciplinary different journals are. 
In our word cloud analysis, we focused on articles cited by different communities, but future work could also examine which papers draw more heavily on mathematics, the life sciences, or both, and compare these patterns to a mathematical biology baseline. In addition, a more in-depth analysis of which modelling approaches 
are valued by different communities would offer important insight into the epistemic priorities shaping the field. 
To extract such information, approaches based on large language models (LLMs) are likely to outperform basic keyword-based methods. 
LLMs could also be used in pipelines for extracting and analysing data on commonly studied cancer types and interventions from article metadata and, alternatively, full article texts.
The potential of LLMs is illustrated by their recent applications in bibliometric research; 
from automating keyword extraction that reflect scientific content \citep{mansour-etal-2025-well} and categorising scientific texts \citep{shaihummel2025_llm}, 
to evaluating research works \citep{evans2024llm}. 
With this rise of LLMs, it becomes increasingly important to ground bibliometric studies in domain knowledge of the studied field, theoretical frameworks, and transparent methodologies. 

\vspace{-.2cm}
\section{Conclusion and outlook}
\vspace{-.2cm}
In this study, we use bibliometric methods to get a bird's-eye view of mathematical oncology research practices. 
We anticipate that our study will encourage further bibliometric research on mathematical oncology and, more broadly, mathematical biology, that quantitatively probes questions that are typically studied qualitatively. 
For instance, a growing collection of literature reviews and perspective articles highlights the importance and potential of mathematical oncology, see e.g., Altrock \citeyearpar{Altrock2015}, Rockne \citeyearpar{Rockne2019_roadmap}, and Kuznetsov \citeyearpar{Kuznetsov2021} et al.. 
In support of these works, our bibliometric study adds data-driven, quantitative evidence of the field's interdisciplinary reach. Further, our study provides insights into how author teams, citation flows, and research topics have changed in the field since the 1960s. 
For us working in mathematical oncology, we can use these insights to inform our research, teaching, organisation, and communication practices. For the good of cancer research, mathematics, and our research community.
\vspace{-.2cm}

\section*{Code and data availability}
\vspace{-.2cm}
The code used in this study is available on the public GitHub repository \url{https://github.com/KiraPugh/Bibliometric_Study_Mathematical_Oncology}. 
Data included in this study are mainly derived from the Web of Science (Clarivate); and partly from Scopus (Elsevier). 
\vspace{-.2cm}

\section*{Acknowledgements}
\vspace{-.2cm}
KP was funded by the Wenner-Gren Stiftelserna/the Wenner-Gren Foundations through a grant to SH (WUP2024-0006).
LG was funded by the Knut and Alice Wallenberg Foundation (KAW 2023.0420), and the Swedish-American Fulbright Commission. 
SH was funded by Wenner-Gren Stiftelserna/the Wenner-Gren Foundations (WGF2022-0044), the Swedish Research Council (project 2024-05621), and the Kjell och M{\"a}rta Beijer Foundation. 
\vspace{-.2cm}

\section*{Author contributions}
\vspace{-.2cm}
Data collection and pre-processing: KP, SS.
Data analysis (implementation) and visualisation: KP.
Data analysis: KP, LG, SH. 
Writing (original draft): KP, LG, SH. 
Writing (editing): all authors.
Conceptualisation: all authors. 
\vspace{-.2cm}

\section*{Declaration of interests}
\vspace{-.2cm}
The authors declare no competing interests.
\vspace{-.2cm}

%\newpage

\begin{spacing}{0.88}
\renewbibmacro{in:}{,}
\section*{References}
\vspace{-.2cm}
\AtNextBibliography{\small}
\printbibliography[heading=none]
\end{spacing}
%TC:endignore
\end{document}

%\setmainfont{Arial}

%\defaultfontfeatures{ Scale=MatchLowercase, Ligatures = TeX }
%\setmainfont{Arial}
%\setsansfont{Arial}
%\setmonofont{Andale Mono}
%\setmathfont{GFSNeohellenicMath.otf}
%\setmathfont[range=up]{Arial}
%\setmathfont[range=it]{Arial Italic}
%\setmathfont[range=bfup]{Arial Bold}
%\setmathfont[range=bfit]{Arial Bold Italic}
%\setmathfont[range=tt]{Andale Mono}
\bigskip
{\Large
\noindent {\bf \emph{Supplementary material to}} A bibliometric study on mathematical oncology: interdisciplinarity, internationality, collaboration and trending topics}
%past and present trends in mathematical oncology
\begin{flushleft}
\bigskip 
Kira Pugh\textsuperscript{$\circledast$}, 
Linnéa Gyllingberg\textsuperscript{},
Stanislav Stratiev\textsuperscript{},
Sara Hamis\textsuperscript{$\circledast$}.

\bigskip

\noindent{\bf $\circledast$ Corresponding authors}: kira.pugh@it.uu.se, sara.hamis@it.uu.se.

\end{flushleft}

\vspace{6cm}
\tableofcontents
\newpage

\section{Journal selection}
We start our journal selection procedure by searching the Web of Science Core Collection (WoSCC) database for articles that match the topic search (TS) keywords below.

\begin{tcolorbox}%[title=Note]
TS = ((cancer* OR tumour* OR tumor* OR oncolog* OR oncogen* OR carcinogen*) AND mathematic*).
\end{tcolorbox}

\noindent Journal-level frequencies of keyword matches are available in Table \ref{tab:journal_selection_QS}, for all journals with 100 or more matches. 
The table also includes journal classification codes based on the Quacquarelli Symonds (QS) system, identifying four journals categorised as mathematics and life sciences journals (underlined) as well as two journals categorised as journals with mathematics, life sciences, and other content (italicised). 
To supplement the journal classification results, we list the self-authored journal descriptions in Table \ref{tab:journal_selection_description}, where we make judgements about which journals focus on mathematical biology (underlined). 
Based on the information in the two tables we classify the following five as our mathematical biology focus journals (in alphabetical order): 
\begin{enumerate}
    \item the Bulletin of Mathematical Biology (BMB),
    \item the Journal of Mathematical Biology (JMB),
     \item the Journal of Theoretical Biology (JTB),
    \item Mathematical Biosciences (MB),
    \item Mathematical Biosciences and Engineering (MBE).
\end{enumerate}

\begin{table}[H]
\caption{\small{ {\bf Journals with more that 100 articles on the WoSCC database
that match the TS keywords in the title, abstract, and/or author keywords.} 
Journals belonging to the QS subject area {\it mathematics}; the QS broad faculty area {\it life sciences and medicine}; and any other QS subject or faculty areas are check-marked.  
}}
\label{tab:journal_selection_QS}
\centering
\small
\resizebox{\textwidth}{!}{
\begin{tabular}{|>{\raggedright\arraybackslash}p{5.2cm}|>{\raggedright\arraybackslash}p{3cm}|>
{\centering\arraybackslash}p{2cm}|>
{\centering\arraybackslash}p{2cm}|>{\centering\arraybackslash}p{2cm}|}
\hline
\textbf{Journal}                                                           & \textbf{Keyword matching article count} & \textbf{Mathematics subject area} & \textbf{Life sciences and medicine faculty area} & \textbf{Other}                         
\\ \hline
\textcolor{red}{\uline {Journal of Theoretical Biology}}    & 455     & \cmark  & \cmark & \textcolor{gray}{}                                                         \\ \hline
PLOS One  & 313   & \textcolor{gray}{}  & \textcolor{gray}{} & \cmark                                                \\ \hline
\textit{\textcolor{blue}{Bulletin of Mathematical Biology}}   & 276                    & \cmark  & \cmark & \cmark                             \\ \hline
Scientific Reports   & 244                                  & \textcolor{gray}{}  & \textcolor{gray}{} & \cmark                                              \\ \hline
Cancer Research     & 239                                          & \textcolor{gray}{}  & \cmark & \textcolor{gray}{}                                                 \\ \hline
\textit{\textcolor{blue}{PLOS Computational Biology}}    & 236                          & \cmark  & \cmark & \cmark   \\ \hline
\textcolor{red}{\uline {Mathematical Biosciences}}           & 222 & \cmark  & \cmark & \textcolor{gray}{}                                                             \\ \hline
\textcolor{red}{\uline {Mathematical Biosciences and Engineering}}     & 160                                      & \cmark  & \cmark & \textcolor{gray}{}                                                       \\ \hline
Medical Physics   & 160                                     & \textcolor{gray}{}  & \cmark & \textcolor{gray}{}                                                  \\ \hline
Physics in Medicine and Biology  & 155                      & \textcolor{gray}{}  & \cmark & \textcolor{gray}{}                                                  \\ \hline
Proceedings of SPIE  & 140                                  & \textcolor{gray}{}  & \textcolor{gray}{} & \textcolor{gray}{}                                                                         \\ \hline
Cancers   & 126  & \textcolor{gray}{}  & \cmark & \textcolor{gray}{}                                                 \\ \hline
Proceedings of the National Academy of Sciences of the United States of America & 121                                          & \textcolor{gray}{}  & \textcolor{gray}{} & \cmark                                               \\ \hline
International Journal of Radiation Oncology Biology Physics                     & 120                                          & \textcolor{gray}{}  & \cmark & \textcolor{gray}{}                                   \\ \hline
\textcolor{red}{\uline {Journal of Mathematical Biology}}    & 119  & \cmark  & \cmark & \textcolor{gray}{}                                                     \\ \hline
\end{tabular}
} %resize
\end{table}

\begin{table}[H]
\caption{{\bf Self-authored descriptions of the journals listed in Table S1.} The descriptions are pasted from the journals' websites. Journals that explicitly highlight both mathematics and biology are underlined. }
\label{tab:journal_selection_description}
{\small
\resizebox{\textwidth}{!}{%
\begin{tabular}{|>{\raggedright\arraybackslash}p{3cm}|>{\raggedright\arraybackslash}p{12.5cm}|}
\hline

\textbf{Journal name}                                                           & \textbf{Summary of aims and scope}                                                                                                                                                                   \\ \hline
\textcolor{red}{\uline {Journal of Theoretical Biology}}                                                  & The Journal of Theoretical Biology is the leading forum for theoretical perspectives that give insight into biological processes. Acceptable papers are those that bear significant importance on the biology per se being presented, and not on the mathematical analysis.                                                                                                                                                                      \\ \hline
PLOS One                                                                        & PLOS ONE welcomes original research submissions from the natural sciences, medical research, engineering, as well as the related social sciences and humanities.                                                                                                                                                                                                                       \\ \hline
\textcolor{red}{\uline {Bulletin of Mathematical Biology}}                                                & As the official journal of the Society for Mathematical Biology, this journal shares research at the biology-mathematics interface. It publishes original research, mathematical biology education, reviews, commentaries, and perspectives.                                                                                                                                                                                                     \\ \hline
Scientific Reports                                                              & We publish original research from all areas of the natural sciences, psychology, medicine and engineering.                                                                                                                                                                                                           \\ \hline
Cancer Research                                                                 & Cancer Research seeks manuscripts that offer conceptual or technological advances leading to basic and translational insights into cancer biology. Manuscripts that focus on convergence science, the bridging of two or more distinct areas of cancer research, are of particular interest.                                                                                                                                                                                      \\ \hline
PLOS Computational Biology                                                      & By making connections through the application of computational methods among disparate areas of biology, PLOS Computational Biology provides substantial new insight into living systems at all scales, from the nano to the macro, and across multiple disciplines, from molecular science, neuroscience and physiology to ecology and population biology.                                                                                                                                                                                                  \\ \hline
\textcolor{red}{\uline {Mathematical Biosciences}}                                                        & Mathematical Biosciences publishes work providing new concepts or new understanding of biological systems using mathematical models, or methodological articles likely to find application to multiple biological systems.                                                                                                                                                                                                                \\ \hline
\textcolor{red}{\uline {Mathematical Biosciences and Engineering}}                                        & MBE focuses on new developments in the fast-growing fields of mathematical biosciences and engineering. Areas covered include most areas of mathematical and computational biology, medicine and engineering with an emphasis on integrative and interdisciplinary research bridging mathematics, biology and engineering.                                                                                                                                                                                                                                           \\ \hline
Medical Physics                                                                 & Medical Physics publishes high-quality articles across radiological imaging, nuclear medicine, radiation oncology, and clinical radiation effects. Topics include new methodologies, imaging or treatment device development, artificial intelligence applications, and clinical or theoretical studies. Radiological imaging covers modalities like CT, MRI, and ultrasound, with a focus on computational analysis such as radiomics and CAD. \\ \hline
Physics in Medicine and Biology                                                 & The novel and impactful development and application of theoretical, computational and experimental physics to medicine, physiology and biology.                                                               \\ \hline
Proceedings of SPIE                                                             & Conference where proceedings are published in the SPIE Digital Library which features more than 600,000 publications in optics and photonics.                                                                                                                                                                                           \\ \hline
Cancers                                                                         & Cancers (ISSN 2072-6694) is an international, peer-reviewed open access journal on oncology.                                                                                                                                                                                                                               \\ \hline
Proceedings of the National Academy of Sciences of the United States of America & PNAS publishes exceptional research across all areas of the biological, physical, and social sciences. Innovation often happens at the margins of disciplines, and we are especially interested in research that crosses disciplinary bounds, answers questions with broad scientific impact, or breaks new ground.                                                                                                                                                                                                                                                                                                                                                                                                                                          \\ \hline
International Journal of Radiation Oncology Biology Physics                     & International Journal of Radiation Oncology - Biology - Physics (IJROBP), known in the field as the Red Journal, publishes original laboratory and clinical investigations related to radiation oncology, radiation biology, medical physics, and both education and health policy as it relates to the field.                                                                                                                                                                                                                                                                                                                                                                                                                                               \\ \hline
\textcolor{red}{\uline {Journal of Mathematical Biology}}                                                 & The Journal of Mathematical Biology (JOMB) focuses on scientific advancements in mathematical modelling and analysis of biological systems.                                                                                                                                      \\ \hline
\end{tabular}
}%end small
}
\end{table}

\newpage
\vspace{-.1cm}
\noindent {\bf Links to journal descriptions are listed below and were accessed on 2025-06-21.}
\vspace{-.2cm}
{\raggedright
\begin{itemize}[leftmargin=*]
    \item Journal of Theoretical Biology: \url{https://www.sciencedirect.com/journal/journal-of-theoretical-biology/about/aims-and-scope}
    \item PLOS One: \url{https://journals.plos.org/plosone/s/journal-information}
    \item Bulletin of Mathematical Biology: \url{https://link.springer.com/journal/11538/aims-and-scope}
    \item Scientific Reports: \url{https://www.nature.com/srep/about}
    \item Cancer Research: \url{https://aacrjournals.org/cancerres/pages/about}
    \item PLOS Computational Biology: \url{https://journals.plos.org/ploscompbiol/s/journal-information}
    \item Mathematical Biosciences: \url{https://www.sciencedirect.com/journal/mathematical-biosciences/about/aims-and-scope}
    \item Mathematical Biosciences and Engineering: \url{https://www.aimspress.com/mbe/news/solo-detail/aimandscope}
    \item Medical Physics: \url{https://aapm.onlinelibrary.wiley.com/hub/journal/24734209/overview}
    \item Physics in Medicine and Biology: \url{https://publishingsupport.iopscience.iop.org/journals/physics-in-medicine-biology/about-physics-medicine-biology}
    \item Proceedings of SPIE: \url{https://www.spiedigitallibrary.org/conference-proceedings-of-spie}
    \item Cancers: \url{https://www.mdpi.com/journal/cancers/about}
    \item Proceedings of the National Academy of Sciences of the United States of America: \url{https://www.pnas.org/author-center}
    \item International Journal of Radiation Oncology Biology Physics: \url{https://www.sciencedirect.com/journal/international-journal-of-radiation-oncology-biology-physics/about/aims-and-scope}
    \item Journal of Mathematical Biology: \url{https://link.springer.com/journal/285/aims-and-scope}
    \vspace{-.2cm}
\end{itemize}
}
\vspace{-.2cm}

\section{Journal classification}
\vspace{-.2cm}
In this study we classify journals as belonging to one of seven categories: 
\vspace{-.1cm}
\begin{enumerate}
{\it 
\item focus journal,
\vspace{-.1cm}
\item mathematics,
\vspace{-.1cm}
\item life sciences, 
\vspace{-.1cm}
\item multidisciplinary, 
\vspace{-.1cm}
\item mathematics and life sciences, 
\vspace{-.1cm}
\item STEM and life sciences, 
\vspace{-.1cm}
\item other.
}
\end{enumerate}
From journal titles in our metadata, we use a previously developed lookup table \citep{himmelstein_scopus} to map journal titles to one or more ASJC codes \citep{elsevierASJC} via Scopus IDs. 
Next, we use the QS scheme \citep{QSranking} which classifies each ASJC code as belonging to one ``broad faculty area" and one ``subject area". 
From a combination of Scopus IDs, ASJC codes, QS broad faculty areas, and QS subject areas we customise the journal classification used in this study, following Algorithm 1.

\begin{algorithm}
\caption{Journal classification.}
\begin{algorithmic}
\State \textbf{Input:} Journal name
\If{Journal \textbf{is} focus journal}
    \State Classify as \textit{focus journal}
\ElsIf{subject area \textbf{is only} mathematics}
    \State Classify as \textit{mathematics}
\ElsIf{broad faculty area \textbf{is only} life sciences and medicine}
    \State Classify as \textit{life sciences}
\ElsIf{ASJC code \textbf{is only} multidisciplinary}
    \State Classify as \textit{multidisciplinary}
\ElsIf{subject area \textbf{includes only} mathematics \textbf{and} a subject area from the broad faculty area life sciences and medicine}
    \State Classify as \textit{mathematics and life sciences}
\ElsIf{broad faculty area \textbf{includes only} (life sciences and medicine) \textbf{and} (natural sciences \textbf{and/or} engineering and technology)}
    \State Classify as \textit{STEM and life sciences}
\Else
    \State Classify as \textit{other}
\EndIf
\State \textbf{Output:} Journal class
\end{algorithmic}
\end{algorithm}

\noindent Codes to map journals to our customised classes are available on the study's GitHub repository.

\section{Journal coverage on the WoSCC database}
In Table \ref{tab:journal_website_vs_woscc} we list the coverage of our focus journals on the WoSCC database.
The coverage is approximated as 
$$
\frac
{\text{\#articles on the WoSCC}}
{\text{\#articles on journal website}} \times 100\%,
$$
where editorials and indexing publications have been omitted.  
The JTB coverage of over 100\% results from retracted publication and addendums being included on the WoSCC.

\vspace{.5cm}

\begin{table}[h]
\caption{\textbf{ 
The WoSCC coverage of the focus journals. 
}
}
\label{tab:journal_website_vs_woscc}
\centering
{\small
\begin{tabular}{|>{\raggedright\arraybackslash}p{3.5cm}|>{\raggedright\arraybackslash}p{3.5cm}|>{\raggedright\arraybackslash}p{3.5cm}|>
{\raggedright\arraybackslash}p{3.5cm}|}
\hline
\textbf{Journal}  & \textbf{\#of articles on the WoSCC} & \textbf{\#of articles on journal website} & \textbf{WoSCC coverage} \\ \hline
Bull. Math. Biol.    & 4102      & 4355                                                                            & 94.19\%        \\ \hline
J. Math. Biol.   & 3689       & 3744                                                                         
& 98.53\%        \\ \hline
J. Theor. Biol.     & 16759      & 16728                                                                         & 100.19\%       \\ \hline
Math. Biosci.     & 5037      & 5329                                                                              & 94.00\%        \\ \hline
Math. Biosci. Eng          &       3946           &     4188            &   94.22\%                                                             \\ \hline
\end{tabular}
}%end small
\end{table}

\newpage
\section{Article classification}
\begin{figure}[h!]
    \centering
    \includegraphics[width=1\linewidth]{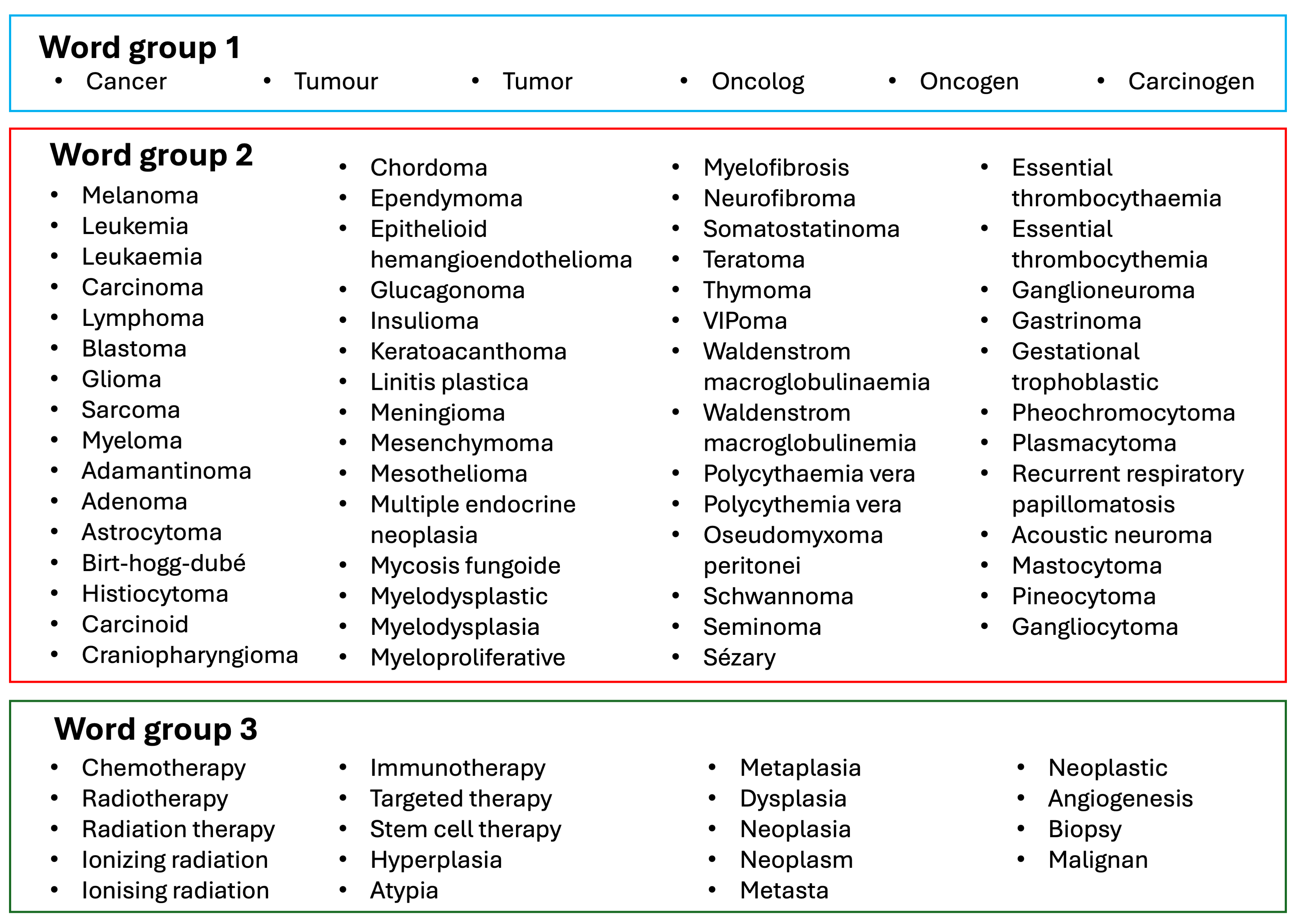}
    \vspace{-.5cm}
    \caption{\textbf{Analysis of cancer-class-keywords.}
    A list of all cancer-class-keywords included in the candidate keyword groups WG1, WG2, and WG3.}
    \label{fig:cancer_keywords}
\end{figure}

\vspace{-.2cm}
In Figure 1 of the main manuscript, we illustrate the procedure of classifying whether an article belongs to the mathematical oncology* or mathematical biology* (excluding mathematical oncology*) dataset. 
This classification is based on keyword matches, and the choice of keywords are based on their predictive performance. 
The list of words used in candidate word groups (WG) 1-3 are shown in Figure \ref{fig:cancer_keywords}, where the words in WG2 include all cancer types that are  listed by the National Cancer Institute \citeyearpar{NCICancerTypes}, Cancer Research UK \citeyearpar{CRUKCancerTypes}, the American Cancer Society \citeyearpar{ACSCancerTypes}, Wikipedia \citeyearpar{WikipediaCancerTypes}, and WebMD \citeyearpar{WebMDCancerAtoZ}; and have subsequently been confirmed as cancer types by the authors.
To assess the accuracy resulting from classification with WG1, WG2$\backslash$WG1, WG3$\backslash$(WG1$\cup$WG2), authors KP and SH reviewed 100 randomly selected articles (20 per focus journal) predicted to be true (in mathematical oncology*), and 100 predicted to be false.
From the confusion matrix in Table \ref{tab:confusion_matrix}, we compute that the accuracy is 0.985 for WG1, 0.925 for WG1\textbackslash WG2, and 0.505 for WG3\textbackslash(WG1$\cup$WG2).
From these results, we decide to use WG1$\cup$WG2 in the cancer-article classification.

\begin{table}[h]
\centering
\begin{tabular}{|l|l|l|}
\hline
& True Positive & True Negative \\ \hline
Predicted Positive & \begin{tabular}[c]{@{}l@{}}WG1: 99\\ WG2\textbackslash WG1: 88\\ WG3\textbackslash (WG1$\cup$WG2): 4\end{tabular} & \begin{tabular}[c]{@{}l@{}}WG1: 1\\ WG2\textbackslash WG1: 12\\ WG3\textbackslash (WG1$\cup$WG2): 96\end{tabular} \\ \hline
Predicted Negative & \begin{tabular}[c]{@{}l@{}}WG1: 2\\ WG2\textbackslash WG1: 3\\ WG3\textbackslash (WG1$\cup$WG2): 3\end{tabular} & \begin{tabular}[c]{@{}l@{}}WG1: 98\\ WG2\textbackslash WG1: 97\\ WG3\textbackslash (WG1$\cup$WG2): 97\end{tabular}                                                                                 \\ \hline
\end{tabular}
\caption{\textbf{Confusion matrix for candidate cancer-class-keywords.}
}
\label{tab:confusion_matrix}
\end{table}

\vspace{-1cm}

\section{Supporting information on interdisciplinarity}

\vspace{-.5cm}

\begin{figure}[h]
    \centering
    \includegraphics[width=0.7\linewidth]{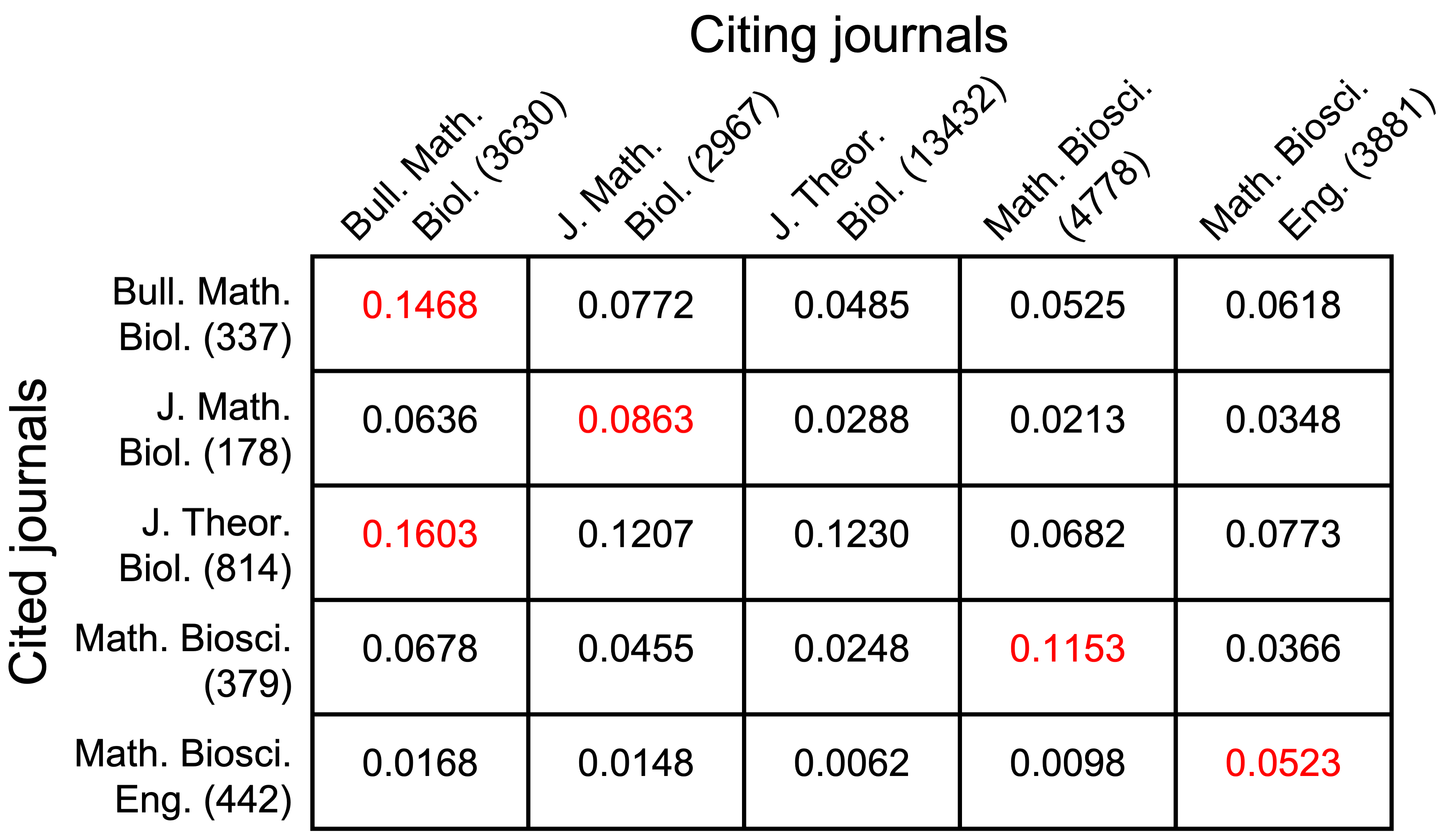}
    \caption{
    \small
    \textbf{Normalised citation matrix.}
    The citation matrix shows how often mathematical oncology* articles in the focus journals (rows) are cited by other focus journals (columns). 
    To account for differences in journal size, the citation counts are normalised by the number of articles in the citing journals (within the mathematical biology* dataset).
    The total number of articles per journal are shown in parentheses. 
     The maximum number of citations per row are red and underlined.}
    \label{fig:normalised_citation_matrix}
\end{figure}

\begin{table}[H]
\caption{\textbf{Top ten journals in each journal category that are cited by and cite mathematical oncology*.} Total article counts are shown in parentheses. } 
\label{tab:top_10_journals}
    \centering
    \small
%\resizebox{\textwidth}{!}{%x
\begin{tabular}{|>{\raggedright\arraybackslash}p{3cm}|>
{\raggedright\arraybackslash}p{5.8cm}|>{\raggedright\arraybackslash}p{5.8cm}|}
    \hline
        \textbf{Category} & \textbf{Journal name (cited by mathematical oncology*)} & \textbf{Journal name (citing mathematical oncology*)} \\ \hline
        \rowcolor{cyan!15}Mathematics and life sciences & Biometrics (148) & Computational and Mathematical Methods in Medicine (300) \\ \hline
        \rowcolor{cyan!15}Mathematics and life sciences & Computational and Mathematical Methods in Medicine (133)  & Theoretical Biology and Medical Modelling (206)  \\ \hline
        \rowcolor{cyan!15}Mathematics and life sciences & Biology Direct (133)  & Biosystems (194)  \\ \hline
        \rowcolor{cyan!15}Mathematics and life sciences & Theoretical Biology and Medical Modelling (114)  & Biology Direct (103)  \\ \hline
        \rowcolor{cyan!15}Mathematics and life sciences & Biosystems (85)  & IET Systems Biology (82) \\ \hline
        \rowcolor{cyan!15}Mathematics and life sciences & Statistics in Medicine (71)  & Statistics in Medicine (70) \\ \hline
        \rowcolor{cyan!15}Mathematics and life sciences & CPT: Pharmacometrics \& Systems Pharmacology (60)  & Biometrics (65) \\ \hline
        \rowcolor{cyan!15}Mathematics and life sciences & IET Systems Biology (40)  & CPT-Pharmacometrics \& Systems Pharmacology (62)  \\ \hline
        \rowcolor{cyan!15}Mathematics and life sciences & Statistical Methods in Medical Research (23)  & Current Bioinformatics (60)  \\ \hline
        \rowcolor{cyan!15}Mathematics and life sciences & Cellular and Molecular Bioengineering (19)  & Cellular and Molecular Bioengineering (52) \\ \hline
        \rowcolor{red!15}Mathematics & Mathematical Models \& Methods in Applied Sciences (393) & Discrete and Continuous Dynamical Systems-Series B (641) \\ \hline
        \rowcolor{red!15}Mathematics & Discrete and Continuous Dynamical Systems-Series B (307) & Mathematical Models \& Methods in Applied Sciences (489) \\ \hline
        \rowcolor{red!15}Mathematics & SIAM Journal on Applied Mathematics (297) & Mathematical Modelling of Natural Phenomena (465)  \\ \hline
        \rowcolor{red!15}Mathematics & Mathematical Modelling of Natural Phenomena (173)  & Communications in Nonlinear Science and Numerical Simulation (314) \\ \hline
    \end{tabular}%
   % }
\end{table}

\begin{table}[H]
\caption*{Table S5.1 continued.}
    \centering
    \small
%\resizebox{\textwidth}{!}{%
\begin{tabular}{|>{\raggedright\arraybackslash}p{3cm}|>
{\raggedright\arraybackslash}p{5.8cm}|>{\raggedright\arraybackslash}p{5.8cm}|}
    \hline
        %\textbf{Category} & \textbf{Journal name (cited by mathematical oncology*)} & \textbf{Journal name (cite mathematical oncology*)} \\ \hline
        \rowcolor{red!15}Mathematics & SIAM Review (129)  & International Journal of Biomathematics (244)  \\ \hline
        \rowcolor{red!15}Mathematics & Journal of Mathematical Analysis and Applications  (114) & Journal of Mathematical Analysis and Applications (212) \\ \hline
        \rowcolor{red!15}Mathematics & Applied Mathematics Letters (108)  & Applied Mathematical Modelling (208) \\ \hline
        \rowcolor{red!15}Mathematics & European Journal of Applied Mathematics (93)  & SIAM Journal on Applied Mathematics (202) \\ \hline
        \rowcolor{red!15}Mathematics & SIAM Journal on Mathematical Analysis (78)  & Journal of Differential Equations (163) \\ \hline
       \rowcolor{red!15} Mathematics & Studies in Applied Mathematics (70)  & AIMS Mathematics (143) \\ \hline
    \rowcolor{green!15}Life sciences & Cancer Research (3371)  & Cancer Research (371) \\ \hline
        \rowcolor{green!15}Life sciences & Cell (1233)  & Cancers (351)  \\ \hline
        \rowcolor{green!15}Life sciences & Nature Reviews Cancer (1032)  & Frontiers in Oncology (298) \\ \hline
        \rowcolor{green!15}Life sciences & British Journal of Cancer (859)  & Physics in Medicine and Biology (207)  \\ \hline
        \rowcolor{green!15}Life sciences & Biophysical Journal (749)  & Biophysical Journal (190)  \\ \hline
        \rowcolor{green!15}Life sciences & Journal of Biological Chemistry (695) & Wiley Interdisciplinary Reviews-Systems Biology and Medicine (171)  \\ \hline
        \rowcolor{green!15}Life sciences & Blood  (626) & Cell Proliferation (154)  \\ \hline
        \rowcolor{green!15}Life sciences & Clinical Cancer Research (558)  & Physical Biology (149)  \\ \hline
        \rowcolor{green!15}Life sciences & New England Journal of Medicine (514) & Frontiers in Immunology (142)  \\ \hline
        \rowcolor{green!15}Life sciences & Journal of Immunology (494) & Seminars in Cancer Biology (126)  \\ \hline
        \rowcolor{Violet!15}Multidisciplinary & Proceedings of the National Academy of Sciences of the United States of America (2280) & PLOS One (1134)  \\ \hline
        \rowcolor{Violet!15}Multidisciplinary & Nature (2170)  & Scientific Reports (692) \\ \hline
        \rowcolor{Violet!15}Multidisciplinary & Science (1490)  & Proceedings of the National Academy of Sciences of the United States of America (191)  \\ \hline
        \rowcolor{Violet!15}Multidisciplinary & PLOS One (1182)  & Royal Society Open Science (166)  \\ \hline
        \rowcolor{Violet!15}Multidisciplinary & Scientific Reports (465) & iScience (112)  \\ \hline
        \rowcolor{Violet!15}Multidisciplinary & Scientific American (62)  & Heliyon (76) \\ \hline
        \rowcolor{Violet!15}Multidisciplinary & Science Advances (36) & Nature (26) \\ \hline
        \rowcolor{Violet!15}Multidisciplinary & Royal Society Open Science (29) & Science Advances (17)  \\ \hline
        \rowcolor{Violet!15}Multidisciplinary & iScience (14) & omptes Rendus de L'Academie Bulgare des Sciences (15)   \\ \hline
        \rowcolor{Violet!15}Multidisciplinary & Chinese Science Bulletin (8)  & Science (14) \\ \hline
        \rowcolor{yellow!15}STEM and life sciences & PLOS Computational Biology (751) & PLOS Computational Biology (879)  \\ \hline
        \rowcolor{yellow!15}STEM and life sciences & Nature Communications (346)  & Journal of the Royal Society Interface (354)  \\ \hline
        \rowcolor{yellow!15}STEM and life sciences & Radiation Research (344)  & Journal of Biological Systems (238)  \\ \hline
        \rowcolor{yellow!15}STEM and life sciences & Journal of the Royal Society Interface (314)  & Computers in Biology and Medicine (226)  \\ \hline
        \rowcolor{yellow!15}STEM and life sciences & International Journal of Radiation Oncology Biology Physics (294)  & International Journal of Molecular Sciences (226) \\ \hline
        \rowcolor{yellow!15}STEM and life sciences & International Journal of Molecular Sciences (179)  & Biomechanics and Modeling in Mechanobiology (174) \\ \hline
        \rowcolor{yellow!15}STEM and life sciences & Nature Biotechnology (168) & BMC Systems Biology (156) \\ \hline
        \rowcolor{yellow!15}STEM and life sciences & Risk Analysis (157)  & Biomedical Signal Processing and Control (121)  \\ \hline
    \end{tabular}%}
\end{table}

\begin{table}[H]
\caption*{Table S5.1 continued.}
    \centering
    \small
%\resizebox{\textwidth}{!}{%
\begin{tabular}{|>{\raggedright\arraybackslash}p{3cm}|>
{\raggedright\arraybackslash}p{5.8cm}|>{\raggedright\arraybackslash}p{5.8cm}|}
    \hline
    \rowcolor{yellow!15}STEM and life sciences & BMC Systems Biology (142)  & International Journal for Numerical Methods in Biomedical Engineering (119) \\ \hline
        \rowcolor{yellow!15}STEM and life sciences & Proceedings of the Royal Society B-Biological Sciences (140)  & NPJ Systems Biology and Spplications (114)  \\ \hline
\rowcolor{black!15}Other & Mathematical and Computer Modelling (523)  & Physical Review E (488)  \\ \hline
        \rowcolor{black!15}Other & Physical Review E (513) & Mathematical and Computer Modelling  (413) \\ \hline
        \rowcolor{black!15}Other & Physical Review Letters (250)  & Chaos Solitons \& Fractals  (392) \\ \hline
        \rowcolor{black!15}Other & Journal of Computational Physics (186) & Mathematical Methods in the Applied Sciences  (348) \\ \hline
        \rowcolor{black!15}Other & Development (130)  & Mathematics (321)  \\ \hline
        \rowcolor{black!15}Other & Nonlinearity (97)  & Nonlinear Analysis-Real World Applications (305)   \\ \hline
        \rowcolor{black!15}Other & Journal of Chemical Physics (96)  & Computers \& Mathematics with Applications (198)   \\ \hline
        \rowcolor{black!15}Other & Journal of the American Statistical Association (83)  & Nonlinearity (197)   \\ \hline
        \rowcolor{black!15}Other & IEEE Access (79) & European Physical Journal Plus (150)   \\ \hline
        \rowcolor{black!15}Other & Multiscale Modeling \& Simulation (79)  & IEEE Access (119) \\ \hline
    \end{tabular}%}
\end{table}

\begin{figure}[!h]
    \centering
    \includegraphics[width=1\linewidth]{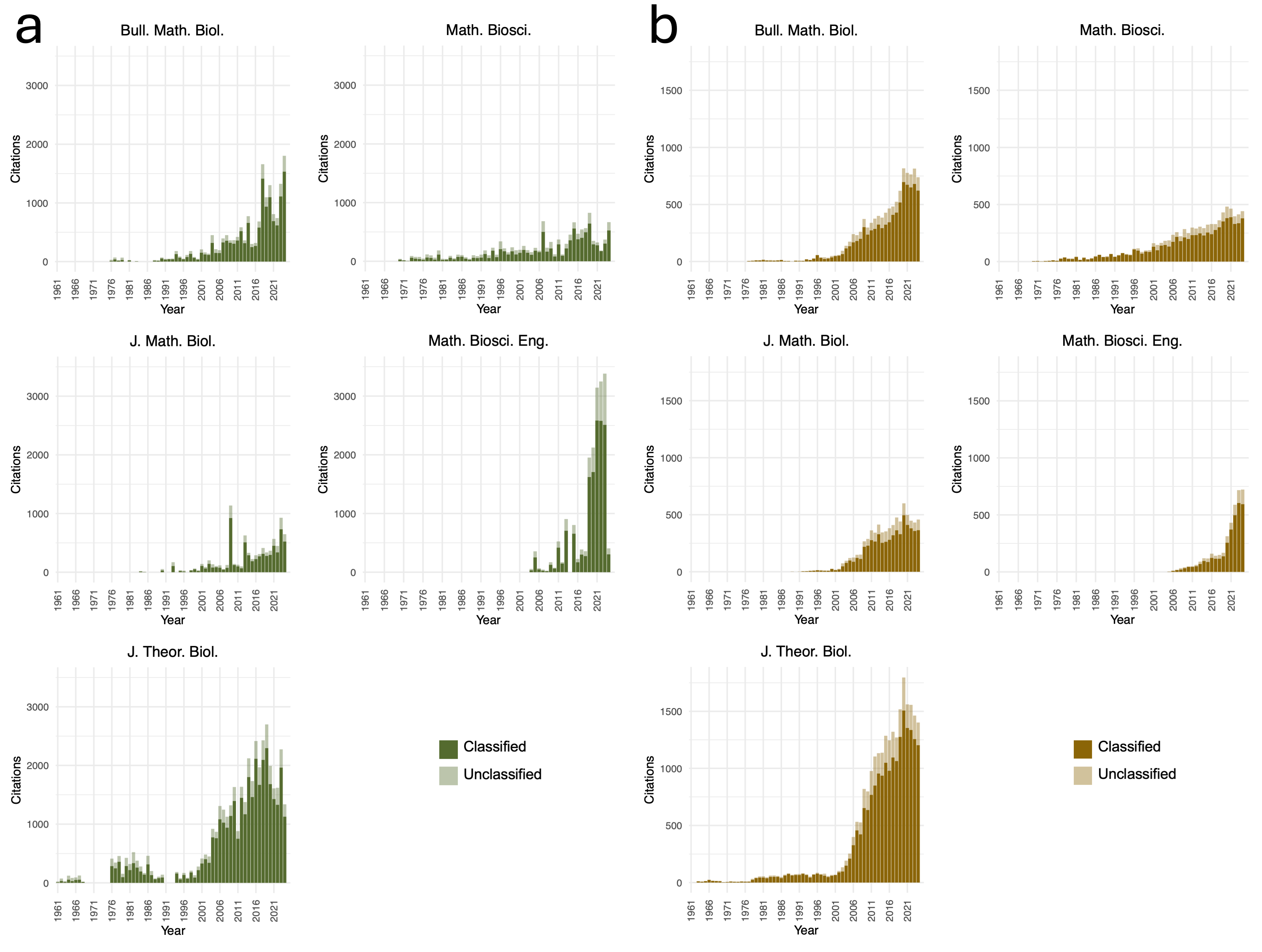}
    \caption{\textbf{Annual counts for articles that are cited by (left) and that cite (bottom) mathematical oncology*.} Counts for both discipline-classified and unclassified journals are shown. }
\label{fig:journal_specific_refernces_citedby}
\end{figure}

\newpage

\section{Supporting information on internationality}

We have affiliation-county metadata for 94.5\% of the articles in the mathematical oncology* dataset, and 87.7\% of the mathematical biology* (excluding mathematical oncology*) dataset .
Of the 2150 mathematical oncology* articles, 118 articles have unspecified country affiliations. 
Among these, 101 articles have no affiliation information available when article metadata is downloaded from WoSCC (9 from BMB, 3 from JMB, 49 from JTB, and 40 from MB), with only 2 published in or after 2000.
The remaining 17 articles, all published before 2000, contain unrecognised country names.
Therefore, per time era the country coverage of the mathematical oncology* dataset is 64.5\% (1961-1999), 99.8\% (2000-2019), and 100\% (2020-2024).

\begin{figure}[H]
    \centering
    \includegraphics[width=1\linewidth]{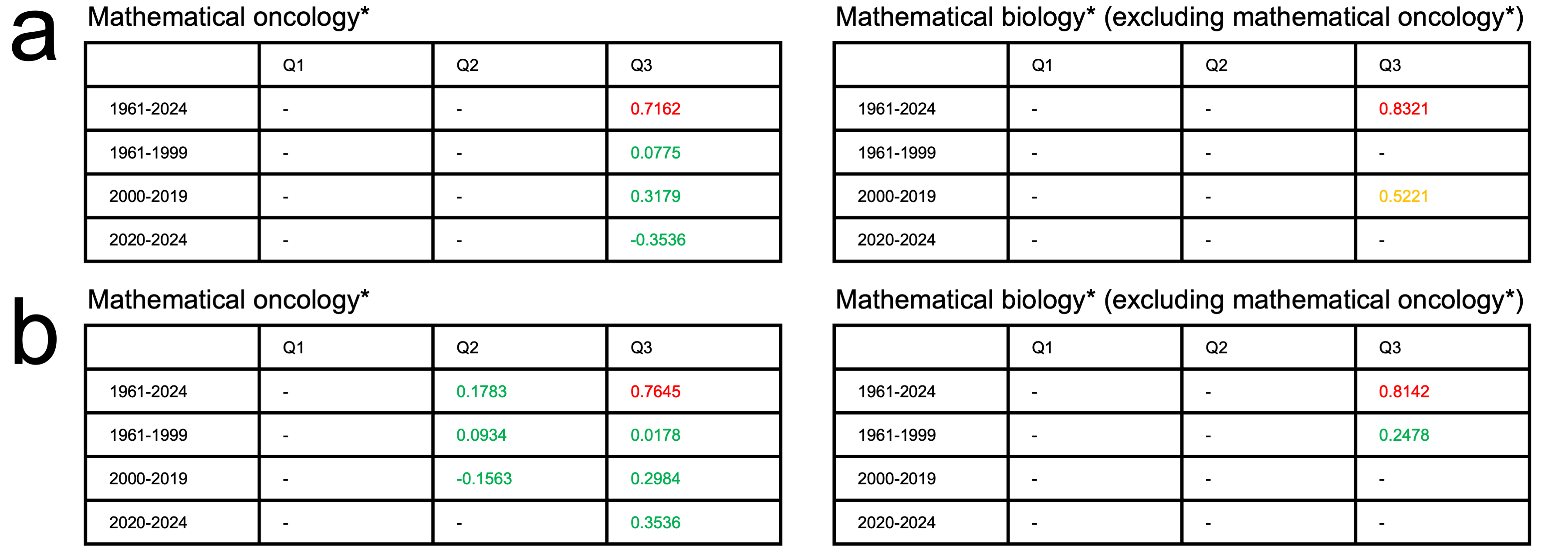}
    \caption{\textbf{Pearson Correlation Coefficients for number of unique author affiliation-countries per article over time.} Values were computed for the lower quartile (Q1), median (Q2), and upper quartile (Q3) over different time periods for (a) all articles in the datasets and (b) those with above-average citations in the corresponding year. Values in red, orange, and green denote strong (0.7 < r $\le$ 1), moderate (0.5 $\le$ r < 0.7), and no or weak (r < 0.5) Pearson correlation coefficients r, respectively. When a tabulated quantity shows no variation across years, the Pearson correlation coefficient is not defined, as is indicated by ‘–’.}
    \label{fig:PCC_international}
\end{figure}

Figure \ref{fig:PCC_international} displays the Pearson correlation coefficients for Q1, Q2, and Q3 over time (1961-2024), as well as for the three time eras we consider in our study, corresponding to the box plots in Figure 3. 
For the three time eras, country-wise article counts are listed in Tables \ref{tab:article_count_era1}-\ref{tab:article_count_era3}; and pairwise collaboration counts are listed in Tables \ref{tab:collab_count_era1}-\ref{tab:collab_count_era3}.

\begin{table}[!ht]
\caption{\bf Number of articles (N) in the mathematical oncology* dataset between 1961 and 1999.}
\footnotesize
    \centering
    \begin{tabular}{|p{2cm}|p{1cm}||p{2cm}|p{1cm}||p{2cm}|p{1cm}||p{2cm}|p{1cm}|}%|l|l||l|l||l|l||l|l|}
    \hline
\rowcolor{orange!15}  \textbf{Country} & \textbf{N} & \textbf{Country} & \textbf{N} & \textbf{Country} & \textbf{N} & \textbf{Country} & \textbf{N} \\ \hline
        USA  & 218 & Sweden   & 7 & Russia   & 4 & Croatia   & 1 \\ \hline
        UK  & 42 & Italy   & 6 & India   & 3 & Denmark   & 1 \\ \hline
        Australia   & 18 & Israel   & 5 & Poland   & 3 & Nigeria   & 1 \\ \hline
        Canada   & 18 & Brazil   & 4 & Austria   & 2 & Norway   & 1 \\ \hline
        Germany   & 15 & Finland   & 4 &Czechia   & 2 & Slovakia   & 1 \\ \hline
        Japan   & 15 & France   & 4 & Bulgaria   & 1 & Sri Lanka   & 1 \\ \hline
        Belgium   & 8 & Netherlands   & 4 & China   & 1 & {\it Unspecified} & 116 \\ \hline
    \end{tabular}
    \label{tab:article_count_era1}
\end{table}

\begin{table}[!ht]
\caption{\bf Number of articles (N) in the mathematical oncology* dataset between 2000 and 2019.}
\footnotesize
    \centering
     \begin{tabular}{|p{2cm}|p{1cm}||p{2cm}|p{1cm}||p{2cm}|p{1cm}||p{2cm}|p{1cm}|}%|l|l||l|l||l|l||l|l|}
    \hline
  \rowcolor{green!15}   \textbf{Country} & \textbf{N} & \textbf{Country} & \textbf{N} & \textbf{Country} & \textbf{N} & \textbf{Country} & \textbf{N} \\ \hline
        USA & 1126 & India & 30 & Austria & 9 & Indonesia & 3 \\ \hline
        United UK & 395 & Netherlands & 28 & Norway & 9 & Kuwait & 3 \\ \hline
        China & 229 & Portugal & 23 & Turkey & 9 & Qatar & 3 \\ \hline
        France & 217 & Russia & 23 & Romania & 8 & Algeria & 2 \\ \hline
        Italy & 168 & Cuba & 16 & Finland & 7 & Bulgaria & 2 \\ \hline
        Canada & 133 & Pakistan & 16 & Saudi Arabia & 6 & Morocco & 2 \\ \hline
        Germany & 116 & Argentina & 15 & Hungary & 5 & Bangladesh & 1 \\ \hline
        Spain & 82 & Belgium & 15 & Singapore & 5 & Botswana & 1 \\ \hline
        Australia & 81 & Switzerland & 15 & Egypt & 4 & Colombia & 1 \\ \hline
        Iran & 61 & Sweden & 12 & Serbia & 4 &Czechia & 1 \\ \hline
        Japan & 52 & Ireland & 11 & Slovakia & 4 & Kenya & 1 \\ \hline
        Poland & 38 & South Korea & 11 & Tunisia & 4 & Lebanon & 1 \\ \hline
        Israel & 37 & Denmark & 10 & Venezuela & 4 & Malaysia & 1 \\ \hline
        Brazil & 32 & Mexico & 10 & Chile & 3 & Eswatini & 1 \\ \hline
        New Zealand & 32 & South Africa & 10 & Greece & 3 & \textit{Unspecified} & 2 \\ \hline
    \end{tabular}
    \label{tab:article_count_era2}
\end{table}

\begin{table}[!ht]
\caption{\bf Number of articles (N) in the mathematical oncology* dataset between 2020 and 2024.}
\footnotesize
    \centering
     \begin{tabular}{|p{2cm}|p{1cm}||p{2cm}|p{1cm}||p{2cm}|p{1cm}||p{2cm}|p{1cm}|}%|l|l||l|l||l|l||l|l|}
    \hline
      \rowcolor{violet!15}  \textbf{Country} & \textbf{N} & \textbf{Country} & \textbf{N} & \textbf{Country} & \textbf{N} & \textbf{Country} & \textbf{N} \\ \hline
        China   & 533 & Turkey   & 10 & Cyprus   & 4 & Tunisia   & 2 \\ \hline
        USA  & 406 & Chile   & 9 & Iraq   & 4 & Azerbaijan   & 1 \\ \hline
        UK  & 116 & Brazil   & 8 & Morocco   & 4 & Belgium   & 1 \\ \hline
        France   & 71 & Indonesia   & 8 & Hungary   & 3 & Cuba   & 1 \\ \hline
        Italy   & 70 & Netherlands   & 8 & Ireland   & 3 & Kazakhstan   & 1 \\ \hline
        Germany   & 52 & New Zealand   & 8 & Israel   & 3 & Luxembourg   & 1 \\ \hline
        Canada   & 51 & Portugal   & 8 & Lesotho   & 3 & Nigeria   & 1 \\ \hline
        Australia   & 49 & South Korea   & 8 & Poland   & 3 & Oman   & 1 \\ \hline
        Spain   & 44 & Greece   & 7 & Slovakia   & 3 & Philippines   & 1 \\ \hline
        India   & 39 & Malaysia   & 7 & Uganda   & 3 & Senegal   & 1 \\ \hline
        Sweden   & 27 & Russia   & 7 & Algeria   & 2 & Serbia   & 1 \\ \hline
        Norway   & 22 & Switzerland   & 7 & Argentina   & 2 & Singapore   & 1 \\ \hline
        Iran   & 20 & Colombia   & 6 & Bangladesh   & 2 & Ukraine   & 1 \\ \hline
        Pakistan   & 18 & Denmark   & 6 & Benin   & 2 & Yemen   & 1 \\ \hline
        Finland   & 17 & Egypt   & 6 & Lebanon   & 2 & Zimbabwe   & 1 \\ \hline
        Saudi Arabia   & 16 & Japan   & 6 & Romania   & 2 & \textit{Unspecified} & 0 \\ \hline
        Mexico   & 12 & Vietnam   & 6 & Rwanda   & 2 & ~ & ~ \\ \hline
        South Africa   & 11 & Austria   & 4 & Thailand   & 2 &  &  \\ \hline
    \end{tabular}
    \label{tab:article_count_era3}
\end{table}

\begin{table}[!ht]
\caption{\bf Number of shared articles (N) in the mathematical oncology* dataset between 1961 and 1999.}
\scriptsize
    \centering
%    \begin{tabular}{|l|l|l||l|l|l||l|l|l|}
     \begin{tabular}{|p{1.8cm}|p{1.8cm}|p{.3cm}||p{1.8cm}|p{1.8cm}|p{.3cm}||p{1.8cm}|p{1.8cm}|p{.3cm}|}
    \hline
        \rowcolor{orange!15} \textbf{Country 1} & \textbf{Country 2} & \textbf{N} & \textbf{Country 1} & \textbf{Country 2} & \textbf{N} & \textbf{Country 1} & \textbf{Country 2} & \textbf{N} \\ \hline
        USA & Canada   & 2 & Finland & Russia   & 1 & USA & Italy   & 1 \\ \hline
        USA & Israel   & 2 & Germany & Russia   & 1 & USA & Japan   & 1 \\ \hline
        USA & Poland   & 2 & Sweden & Norway   & 1 & USA & Russia   & 1 \\ \hline
        USA & Sweden   & 2 & USA & Australia   & 1 & ~ & ~ & ~ \\ \hline
    \end{tabular}
    \label{tab:collab_count_era1}
\end{table}

\begin{table}[!ht]
\caption{\bf Number of shared articles (N) in the mathematical oncology* dataset between 2000 and 2019.}
\scriptsize
    \centering
    %\begin{tabular}{|l|l|l||l|l|l||l|l|l|}
     \begin{tabular}{|p{1.8cm}|p{1.8cm}|p{.3cm}||p{1.8cm}|p{1.8cm}|p{.3cm}||p{1.8cm}|p{1.8cm}|p{.3cm}|}
    \hline
        \rowcolor{green!15} \textbf{Country 1} & \textbf{Country 2} & \textbf{N} & \textbf{Country 1} & \textbf{Country 2} & \textbf{N} & \textbf{Country 1} & \textbf{Country 2} & \textbf{N} \\ \hline
        USA   & UK   & 41 & Poland   & Israel   & 2 & Italy   & Venezuela   & 1 \\ \hline
        USA   & China   & 34 & Spain   & Brazil   & 2 & Japan   & Austria   & 1 \\ \hline
        USA   & France   & 18 & UK   & Austria   & 2 & Japan   & Ireland   & 1 \\ \hline
        USA   & Germany   & 16 & UK   & Portugal   & 2 & Netherlands   & Austria   & 1 \\ \hline
        USA   & Canada   & 15 & USA   & Argentina   & 2 & Netherlands   & Finland   & 1 \\ \hline
        UK   & Australia   & 13 & USA   & Pakistan   & 2 & Netherlands   & Sweden   & 1 \\ \hline
        UK   & France   & 13 & USA   & Singapore   & 2 & New Zealand   & Singapore   & 1 \\ \hline
        UK   & Germany   & 13 & USA   & Switzerland   & 2 & Pakistan   & Korea   & 1 \\ \hline
        UK   & Canada   & 10 & Argentina   & Belgium   & 1 & Pakistan   & Saudi Arabia   & 1 \\ \hline
        Italy   & Spain   & 9 & Australia   & India   & 1 & Poland   & Egypt   & 1 \\ \hline
        USA   & Italy   & 9 & Australia   & Mexico   & 1 & Poland   & South Africa   & 1 \\ \hline
        USA   & Japan   & 9 & Australia   & Netherlands   & 1 & Poland   & Switzerland   & 1 \\ \hline
        UK   & Italy   & 7 & Australia   & New Zealand   & 1 & Portugal   & Argentina   & 1 \\ \hline
        France   & Italy   & 6 & Belgium   & Austria   & 1 & Portugal   & Belgium   & 1 \\ \hline
        USA   & Australia   & 6 & Belgium   & Greece   & 1 & Portugal   & Cuba   & 1 \\ \hline
        USA   & India   & 6 & Belgium   & Ireland   & 1 & Portugal   & Mexico   & 1 \\ \hline
        USA   & Poland   & 6 & Brazil   & Cuba   & 1 & Portugal   & South Africa   & 1 \\ \hline
        USA   & Russia   & 6 & Brazil   & Mexico   & 1 & Russia   & Mexico   & 1 \\ \hline
        USA   & Spain   & 6 & Brazil   & Portugal   & 1 & Saudi Arabia   & Egypt   & 1 \\ \hline
        Germany   & Russia   & 5 & Brazil   & Switzerland   & 1 & South Africa   & Bulgaria   & 1 \\ \hline
        Italy   & Germany   & 5 & Canada   & Denmark   & 1 & South Africa   & Swaziland   & 1 \\ \hline
        UK   & Ireland   & 5 & Canada   & Ireland   & 1 & Spain   & Argentina   & 1 \\ \hline
        UK   & Spain   & 5 & Canada   & Israel   & 1 & Spain   & Australia   & 1 \\ \hline
        USA   & Israel   & 5 & Canada   & Japan   & 1 & Spain   & Bulgaria   & 1 \\ \hline
        USA   & Netherlands   & 5 & Canada   & Netherlands   & 1 & Spain   & Cuba   & 1 \\ \hline
        Australia   & Korea   & 4 & Canada   & UAE   & 1 & Spain   & Norway   & 1 \\ \hline
        Canada   & Australia   & 4 & Chile   & Colombia   & 1 & Spain   & Poland   & 1 \\ \hline
        China   & Japan   & 4 & China   & Austria   & 1 & Spain   & Russia   & 1 \\ \hline
        France   & Germany   & 4 & China   & Egypt   & 1 & Spain   & Serbia   & 1 \\ \hline
        UK   & China   & 4 & China   & Germany   & 1 & Spain   & South Africa   & 1 \\ \hline
        USA   & Brazil   & 4 & China   & Italy   & 1 & Spain   & Switzerland   & 1 \\ \hline
        USA   & Denmark   & 4 & China   & Kenya   & 1 & Sweden   & Finland   & 1 \\ \hline
        USA   & Iran   & 4 & China   & New Zealand   & 1 & UK   & Argentina   & 1 \\ \hline
        USA   & Korea   & 4 & China   & Spain   & 1 & UK   & Bangladesh   & 1 \\ \hline
        USA   & Saudi Arabia   & 4 & Cuba   & Mexico   & 1 & UK   & Belgium   & 1 \\ \hline
        USA   & South Africa   & 4 & France   & Argentina   & 1 & UK   & Brazil   & 1 \\ \hline
        Australia   & Poland   & 3 & France   & Brazil   & 1 & UK   & Finland   & 1 \\ \hline
        France   & Israel   & 3 & France   & Canada   & 1 & UK   & Greece   & 1 \\ \hline
        France   & Spain   & 3 & France   & Lebanon   & 1 & UK   & Japan   & 1 \\ \hline
        France   & Switzerland   & 3 & France   & Netherlands   & 1 & UK   & Netherlands   & 1 \\ \hline
        Germany   & Hungary   & 3 & France   & Poland   & 1 & UK   & New Zealand   & 1 \\ \hline
        Germany   & Poland   & 3 & France   & Russia   & 1 & UK   & Poland   & 1 \\ \hline
        Germany   & Spain   & 3 & France   & Saudi Arabia   & 1 & UK   & Qatar   & 1 \\ \hline
        Spain   & Mexico   & 3 & France   & South Africa   & 1 & UK   & Russia   & 1 \\ \hline
        Spain   & Portugal   & 3 & Germany   & Australia   & 1 & UK   & Saudi Arabia   & 1 \\ \hline
        UK   & Israel   & 3 & Germany   & Austria   & 1 & UK   & South Africa   & 1 \\ \hline
        USA   & Austria   & 3 & Germany   & Brazil   & 1 & UK   & Switzerland   & 1 \\ \hline
        USA   & Portugal   & 3 & Germany   & Bulgaria   & 1 & UK   & Turkey   & 1 \\ \hline
        USA   & Qatar   & 3 & Germany   & Greece   & 1 & UK   & UAE   & 1 \\ \hline
        Australia   & Romania   & 2 & Germany   & New Zealand   & 1 & USA   & Bangladesh   & 1 \\ \hline
        Canada   & Germany   & 2 & Germany   & Portugal   & 1 & USA   & Botswana   & 1 \\ \hline
        China   & Australia   & 2 & Germany   & South Africa   & 1 & USA   & Egypt   & 1 \\ \hline
        China   & Canada   & 2 & Iran   & Sweden   & 1 & USA   & Hungary   & 1 \\ \hline
        China   & Saudi Arabia   & 2 & Ireland   & Austria   & 1 & USA   & Indonesia   & 1 \\ \hline
        China   & Singapore   & 2 & Israel   & Mexico   & 1 & USA   & Malaysia   & 1 \\ \hline
        France   & Australia   & 2 & Israel   & Sweden   & 1 & USA   & Mexico   & 1 \\ \hline
        France   & Austria   & 2 & Italy   & Argentina   & 1 & USA   & New Zealand   & 1 \\ \hline
        France   & Japan   & 2 & Italy   & Australia   & 1 & USA   & Romania   & 1 \\ \hline
        France   & Qatar   & 2 & Italy   & Austria   & 1 & USA   & Turkey   & 1 \\ \hline
        Germany   & Switzerland   & 2 & Italy   & Brazil   & 1 & USA   & UAE   & 1 \\ \hline
        Italy   & Canada   & 2 & Italy   & Cuba   & 1 & USA   & Venezuela   & 1 \\ \hline
        Italy   & Mexico   & 2 & Italy   & New Zealand   & 1 & ~ & ~ & ~ \\ \hline
        Italy   & Netherlands   & 2 & Italy   & Switzerland   & 1 & ~ & ~ & ~ \\ \hline
    \end{tabular}
    \label{tab:collab_count_era2}
\end{table}

\begin{table}[!ht]
\caption{\bf Number of shared articles (N) in the mathematical oncology* dataset between 2020 and 2024.}
\scriptsize
    \centering
    %\begin{tabular}{|l|l|l||l|l|l||l|l|l|}
         \begin{tabular}{|p{1.8cm}|p{1.8cm}|p{.3cm}||p{1.8cm}|p{1.8cm}|p{.3cm}||p{1.8cm}|p{1.8cm}|p{.3cm}|}
    \hline
        \rowcolor{violet!15} \textbf{Country 1} & \textbf{Country 2} & \textbf{N} & \textbf{Country 1} & \textbf{Country 2} & \textbf{N} & \textbf{Country 1} & \textbf{Country 2} & \textbf{N} \\ \hline
        USA & UK & 15 & Canada & Saudi Arabia & 1 & Italy & Turkey & 1 \\ \hline
        China & USA & 11 & Canada & South Africa & 1 & Korea & Hungary & 1 \\ \hline
        USA & Canada & 10 & Canada & Thailand & 1 & Korea & Lebanon & 1 \\ \hline
        UK & France & 9 & Canada & Turkey & 1 & Korea & Slovakia & 1 \\ \hline
        UK & Italy & 8 & Canada & Zimbabwe & 1 & Korea & UAE & 1 \\ \hline
        UK & Australia & 7 & Chile & Brazil & 1 & Morocco & Tunisia & 1 \\ \hline
        China & Australia & 5 & Chile & Luxembourg & 1 & Netherlands & Belgium & 1 \\ \hline
        China & UK & 5 & China & Algeria & 1 & Netherlands & Israel & 1 \\ \hline
        France & Italy & 5 & China & Azerbaijan & 1 & Norway & Austria & 1 \\ \hline
        Italy & Spain & 5 & China & Bangladesh & 1 & Norway & Korea & 1 \\ \hline
        China & Pakistan & 4 & China & France & 1 & Norway & Lebanon & 1 \\ \hline
        Pakistan & Saudi Arabia & 4 & China & Nigeria & 1 & Norway & Portugal & 1 \\ \hline
        UK & Spain & 4 & China & Philippines & 1 & Norway & UAE & 1 \\ \hline
        UK & Switzerland & 4 & China & Rwanda & 1 & Pakistan & Azerbaijan & 1 \\ \hline
        USA & France & 4 & China & Saudi Arabia & 1 & Pakistan & Egypt & 1 \\ \hline
        USA & Italy & 4 & China & Singapore & 1 & Pakistan & Hungary & 1 \\ \hline
        USA & Norway & 4 & China & South Africa & 1 & Pakistan & Korea & 1 \\ \hline
        USA & Spain & 4 & China & Spain & 1 & Pakistan & Slovakia & 1 \\ \hline
        Australia & Saudi Arabia & 3 & China & Thailand & 1 & Pakistan & Thailand & 1 \\ \hline
        Canada & Iran & 3 & China & Yemen & 1 & Russia & Benin & 1 \\ \hline
        China & Canada & 3 & Colombia & Zimbabwe & 1 & Russia & Morocco & 1 \\ \hline
        Saudi Arabia & Malaysia & 3 & Egypt & Algeria & 1 & Rwanda & Nigeria & 1 \\ \hline
        South Africa & Lesotho & 3 & Egypt & Yemen & 1 & Saudi Arabia & Egypt & 1 \\ \hline
        UK & Germany & 3 & Finland & Netherlands & 1 & Saudi Arabia & Hungary & 1 \\ \hline
        UK & Saudi Arabia & 3 & France & Chile & 1 & Saudi Arabia & Indonesia & 1 \\ \hline
        USA & Australia & 3 & France & Cyprus & 1 & Saudi Arabia & Iraq & 1 \\ \hline
        USA & Lesotho & 3 & France & Luxembourg & 1 & Saudi Arabia & Korea & 1 \\ \hline
        USA & South Africa & 3 & France & Netherlands & 1 & Saudi Arabia & Slovakia & 1 \\ \hline
        USA & Switzerland & 3 & France & Poland & 1 & South Africa & Algeria & 1 \\ \hline
        Australia & India & 2 & France & Senegal & 1 & South Africa & Colombia & 1 \\ \hline
        Australia & Korea & 2 & France & South Africa & 1 & South Africa & Egypt & 1 \\ \hline
        Australia & Malaysia & 2 & France & Sweden & 1 & South Africa & Yemen & 1 \\ \hline
        China & Egypt & 2 & France & Tunisia & 1 & South Africa & Zimbabwe & 1 \\ \hline
        China & Germany & 2 & France & Zimbabwe & 1 & Spain & Mexico & 1 \\ \hline
        China & Iraq & 2 & Germany & Canada & 1 & Spain & Serbia & 1 \\ \hline
        China & Italy & 2 & Germany & Denmark & 1 & Sweden & Austria & 1 \\ \hline
        China & Japan & 2 & Germany & Iran & 1 & Sweden & Finland & 1 \\ \hline
        China & Norway & 2 & Germany & Israel & 1 & Sweden & Norway & 1 \\ \hline
        Finland & Austria & 2 & Germany & Netherlands & 1 & Sweden & Russia & 1 \\ \hline
        France & Canada & 2 & Germany & New Zealand & 1 & Thailand & Azerbaijan & 1 \\ \hline
        France & Colombia & 2 & Germany & Pakistan & 1 & Turkey & Japan & 1 \\ \hline
        France & Morocco & 2 & Germany & Spain & 1 & Turkey & Morocco & 1 \\ \hline
        France & Russia & 2 & Germany & Turkey & 1 & UAE & Lebanon & 1 \\ \hline
        France & Spain & 2 & Hungary & Slovakia & 1 & UK & Brazil & 1 \\ \hline
        Italy & Australia & 2 & India & Korea & 1 & UK & Canada & 1 \\ \hline
        Italy & Brazil & 2 & India & Lebanon & 1 & UK & Cyprus & 1 \\ \hline
        Italy & Canada & 2 & India & Malaysia & 1 & UK & Finland & 1 \\ \hline
        Norway & Finland & 2 & India & Norway & 1 & UK & Greece & 1 \\ \hline
        Pakistan & Iraq & 2 & India & Pakistan & 1 & UK & India & 1 \\ \hline
        Spain & Brazil & 2 & India & Saudi Arabia & 1 & UK & Nigeria & 1 \\ \hline
        Sweden & Morocco & 2 & India & UAE & 1 & UK & Norway & 1 \\ \hline
        UK & Ireland & 2 & Indonesia & Malaysia & 1 & UK & Oman & 1 \\ \hline
        UK & Netherlands & 2 & Indonesia & Ukraine & 1 & UK & Rwanda & 1 \\ \hline
        USA & Germany & 2 & Iran & Egypt & 1 & UK & Sweden & 1 \\ \hline
        USA & Netherlands & 2 & Iran & Finland & 1 & USA & Algeria & 1 \\ \hline
        USA & Russia & 2 & Iran & Pakistan & 1 & USA & Brazil & 1 \\ \hline
        USA & Sweden & 2 & Iran & Saudi Arabia & 1 & USA & Finland & 1 \\ \hline
        Algeria & Thailand & 1 & Iraq & Azerbaijan & 1 & USA & Greece & 1 \\ \hline
        Algeria & Yemen & 1 & Iraq & Thailand & 1 & USA & Kazakhstan & 1 \\ \hline
        Australia & Indonesia & 1 & Israel & Belgium & 1 & USA & Korea & 1 \\ \hline
        Australia & New Zealand & 1 & Italy & Austria & 1 & USA & Mexico & 1 \\ \hline
        Canada & Australia & 1 & Italy & Azerbaijan & 1 & USA & Morocco & 1 \\ \hline
        Canada & Azerbaijan & 1 & Italy & India & 1 & USA & New Zealand & 1 \\ \hline
        Canada & Colombia & 1 & Italy & Iraq & 1 & USA & Pakistan & 1 \\ \hline
        Canada & Iraq & 1 & Italy & Morocco & 1 & USA & Poland & 1 \\ \hline
        Canada & Korea & 1 & Italy & New Zealand & 1 & USA & Senegal & 1 \\ \hline
        Canada & Malaysia & 1 & Italy & Pakistan & 1 & USA & Thailand & 1 \\ \hline
        Canada & New Zealand & 1 & Italy & Sweden & 1 & ~ & ~ & ~ \\ \hline
        Canada & Pakistan & 1 & Italy & Thailand & 1 & ~ & ~ & ~ \\ \hline
    \end{tabular}
    \label{tab:collab_count_era3}
\end{table}

\section{Supporting information on collaboration}

\begin{figure}[h]
    \centering
    \includegraphics[width=\linewidth]{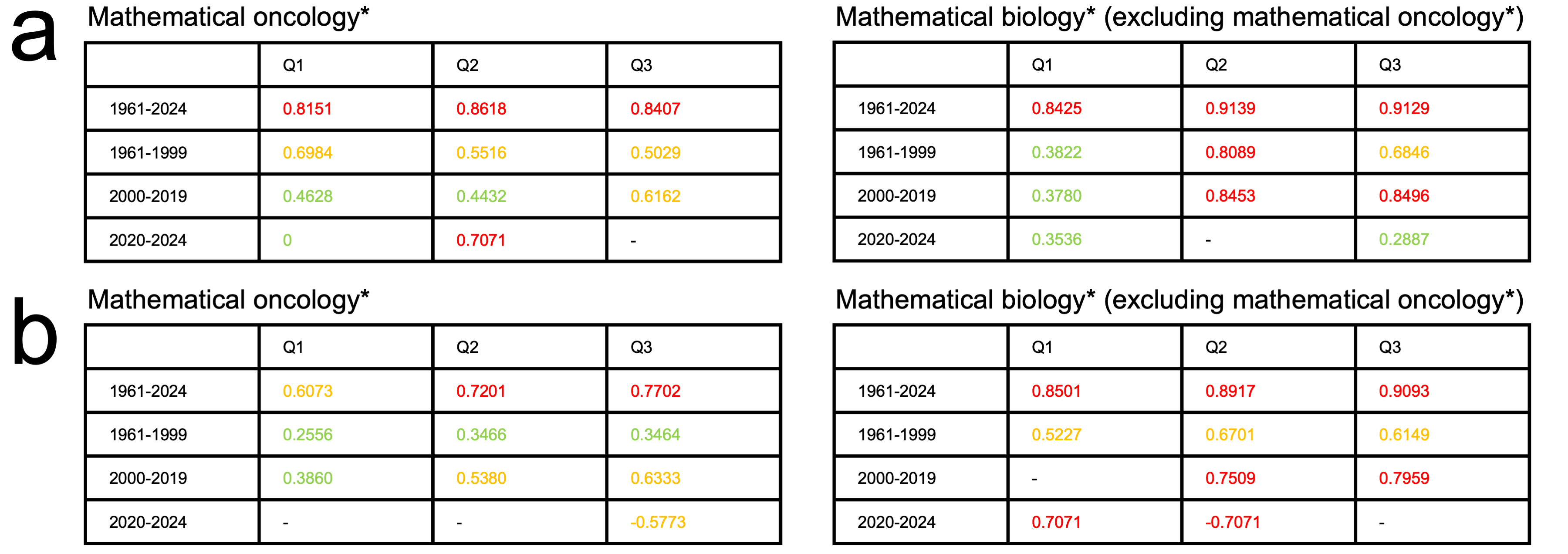}
    \caption{\textbf{Pearson Correlation Coefficients for the number of authors per article over time.} Values were computed for the lower quartile (Q1), median (Q2), and upper quartile (Q3) over different time periods for (a) all articles in the datasets and (b) those with above-average citations in the corresponding year. Values in red, orange, and green denote strong (0.7 < r $\le$ 1), moderate (0.5 $\le$ r < 0.7), and no or weak (r < 0.5) Pearson correlation coefficients r, respectively. When a tabulated quantity shows no variation across years, the Pearson correlation coefficient is not defined, as is indicated by ‘–’.}
    \label{fig:PCC_collaboration}
\end{figure}

Figure \ref{fig:PCC_collaboration} displays the Pearson correlation coefficients for Q1, Q2, and Q3 for all box plots in Figure 4 in the main manuscript.

\section{Supporting information on trending topics}
\begin{figure}
    \centering
    \includegraphics[width=1\linewidth]{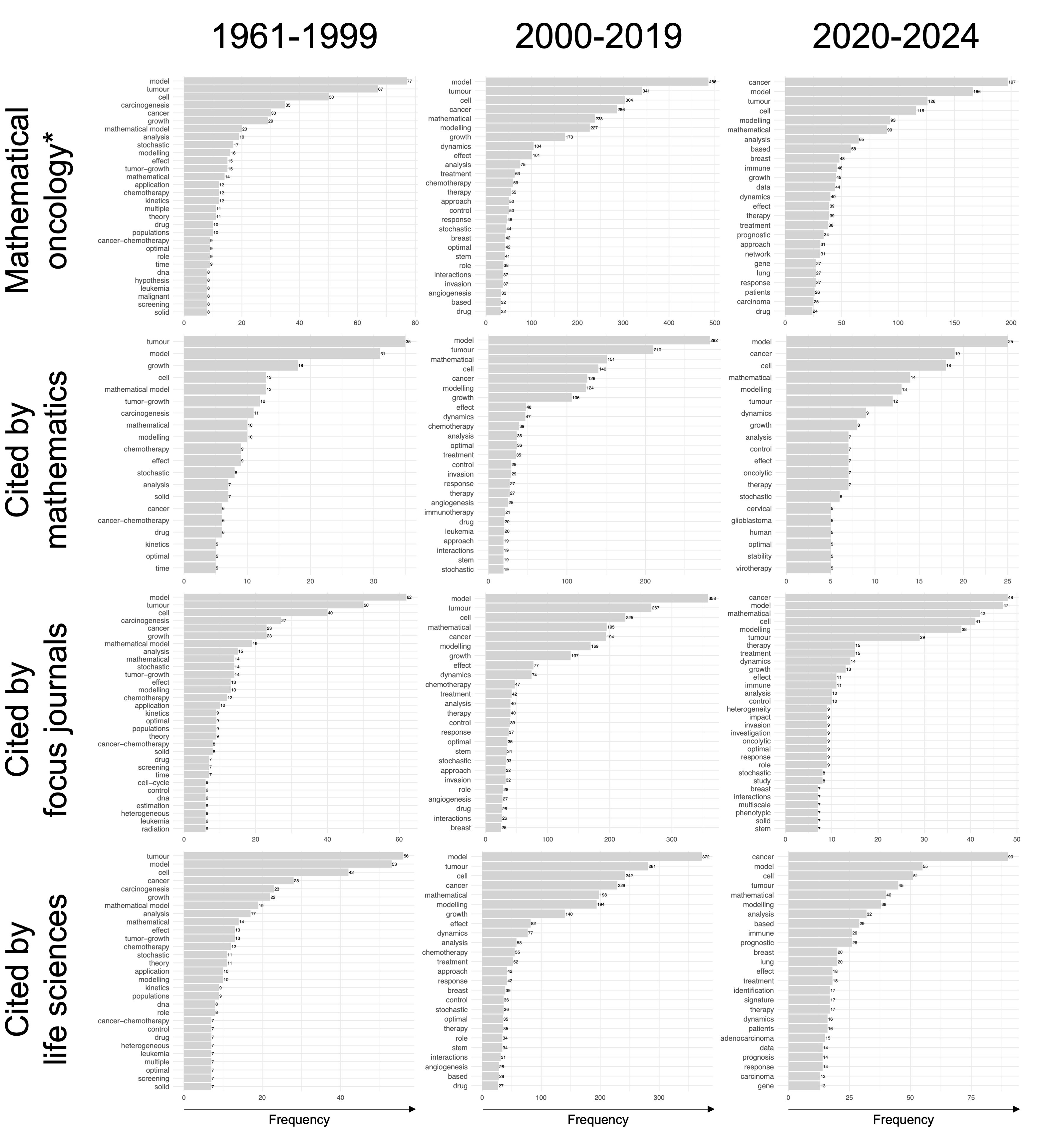}
    \caption{{\bf Trending topics in mathematical oncology analysed through title word frequency.}
     The bar charts contain the 25 most frequent terms, with frequencies next to the bars. 
     Data for all mathematical oncology* titles, per time period, are shown in the top row. 
     Data for articles with at least one citation in mathematics, focus, and life science journals are shown in the three rows below, respectively. }
    \label{fig:barchart_titles}
\end{figure}

\begin{figure}
    \centering
    \includegraphics[width=1\linewidth]{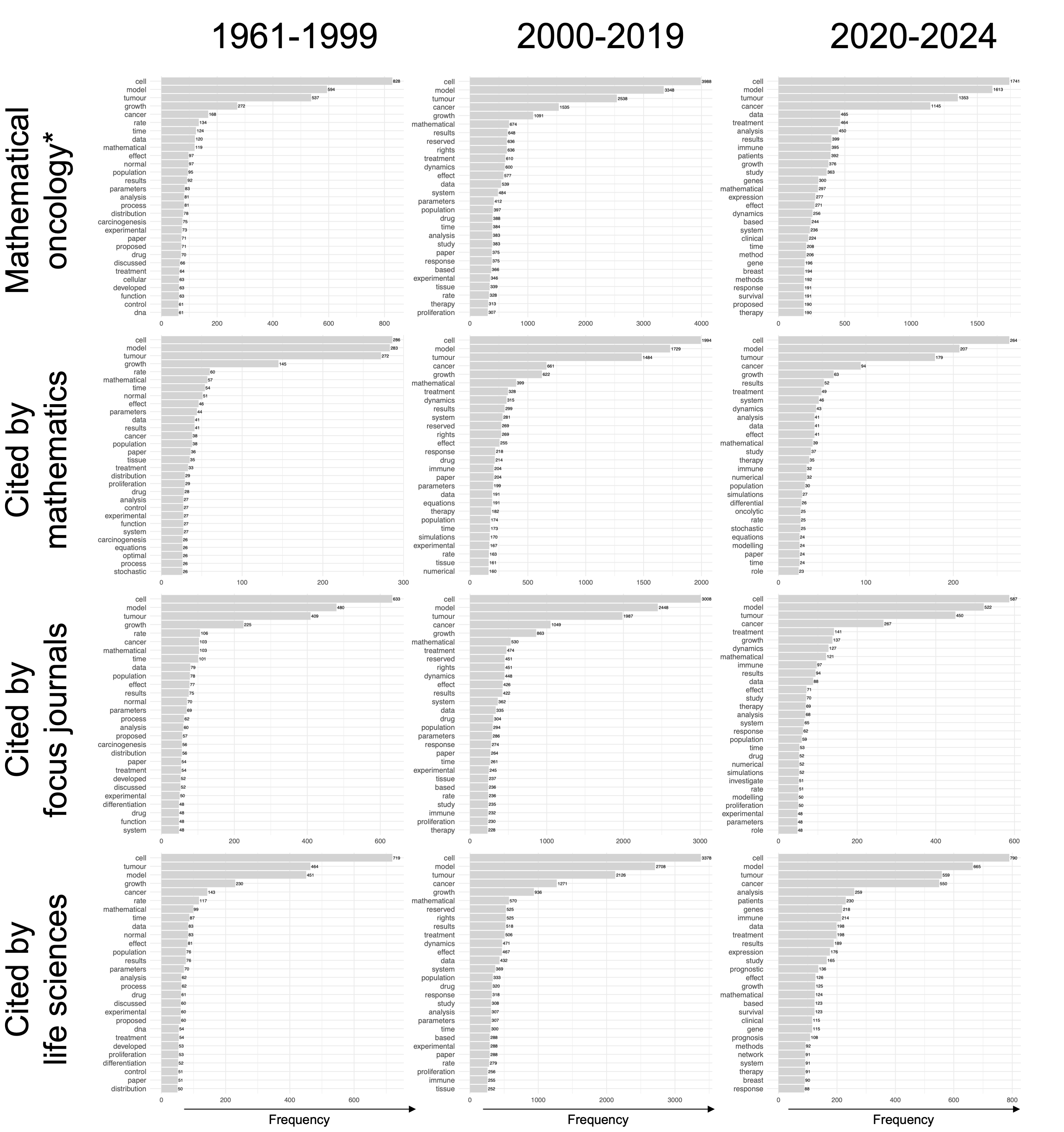}
    \caption{{\bf Trending topics in mathematical oncology analysed through abstract word frequency.}
     The bar charts contain the 25 most frequent terms, with frequencies next to the bars. 
     Data for all mathematical oncology* abstract, per time period, are shown in the top row. 
     Data for articles with at least one citation in mathematics, focus, and life science journals are shown in the three rows below, respectively. }
    \label{fig:barchart_abstract}
\end{figure}

The frequency of the word clouds presented in Figures 5 and 6 in the main manuscript are generated with the R-package biblioshiny \citep{aria2017bibliometrix}. 
Synonyms are combined in the word clouds as shown in Table \ref{tab:word_cloud_synonyms} and for the abstract word clouds we remove the words \texttt{cell}, \texttt{model}, \texttt{tumour} according to a similarity pruning. 
For each word cloud, the 25 most frequent words are included; as long as they appear at least 5 times.  
In cases where multiple words (with frequency of at least 5) are tied for 25th place, they are all included. 
Complementary to the word clouds, bar charts displaying the frequency of the top 25 words in article titles and abstracts are shown in Figures \ref{fig:barchart_titles} and \ref{fig:barchart_abstract} respectively. In the study's GitHub repository, word cloud frequencies are available in numeric form in the file \texttt{word\_cloud\_data.csv}.

\begin{table}[h]
\caption{\textbf{A list of synonyms used when generating the word clouds.}}
\label{tab:word_cloud_synonyms}
\centering
\begin{tabular}{|p{3.5cm}|p{11.5cm}|}
\hline
\textbf{Term}               & \textbf{Synonyms}     
\\ \hline
Mathematical model               & Mathematical-model; Mathematical models; Mathematical-models                                                                                                 \\ \hline
Tumour               & Tumor, Tumours, Tumors                                                                                                   \\ \hline
Cell               & Cells                                                                                                  \\ \hline
Modelling          & Modeling                                                                                               \\ \hline
Model              & Models                                                                                                 \\ \hline
Effect             & Effects                                                                                                \\ \hline
\end{tabular}
\end{table}

\section{Code files}
The code files used in this study are available on the public GitHub repository \url{https://github.com/KiraPugh/Bibliometric_Study_Mathematical_Oncology}. 
Instructions on how to run the code files are available in the repository's README file.

\begin{spacing}{0.9}
\renewbibmacro{in:}{,}
\section*{References}
\AtNextBibliography{\small}
\printbibliography[heading=none]
\end{spacing}
%TC:endignore